\documentclass[12pt,reqno]{amsart}
\usepackage{amsmath,amssymb,amsfonts,amsthm}
\usepackage[mathscr]{eucal}
\usepackage[all]{xy}
\usepackage{hyperref}
\usepackage{graphicx}
\usepackage{setspace}
\usepackage{ytableau}
\usepackage{float}
\allowdisplaybreaks

\author{S.L.~Lyakhovich, N.A.~Sinelnikov}

\address{Laboratory of Theoretical and Mathematical Physics, Physics Faculty, Tomsk State University, Novosobornaya Square 1, Tomsk 634050, Russia}

\email{sll@phys.tsu.ru; sinelnikov@phys.tsu.ru}

\title{Gauge symmetry and partially Lagrangian systems}
\begin{document}

\begin{abstract}
We consider a classical field theory whose equations of motion follow from the least action principle, but the class of admissible trajectories is restricted by differential equations. The key element of the proposed construction is the complete gauge symmetry of these additional equations. The unfree variation of the trajectories reduces to the infinitesimal gauge symmetry transformation of the equations restricting the trajectories. We explicitly derive the equations that follow from the requirement that this gauge variation of the action vanishes. The system of equations for conditional extrema is not a Lagrangian system as such, but it admits an equivalent Hamiltonian formulation with a non-canonical Poisson bracket. The bracket is degenerate, in general. Alternatively, the equations restricting the dynamics could be added to the action with Lagrange multipliers with unrestricted variation of the original variables. In this case, we would arrive at the Lagrangian equations for the original variables involving Lagrange multipliers and for Lagrange multipliers themselves. In general, these two methods are not equivalent because the multipliers can bring extra degrees of freedom compared to the case of equations derived by unfree variation of the action. We illustrate the general method with two examples. The first example is a particle in a central field with varying trajectories restricted by the equation of conservation of angular momentum. The phase space acquires one more dimension, and there is an extra conserved quantity $K$ which is responsible for the precession of trajectories. $K=0$ corresponds to the trajectories of usual Lagrangian dynamics. The second example is linearized gravity with the Einstein-Hilbert action, and the class of varying fields is restricted by the linearized Nordstr\"om equation. This conditional extrema problem is shown to lead to the linearized Cotton gravity equations.
\end{abstract}
\maketitle

\section{Introduction}
In recent years, there has been growing interest in field theories whose classical dynamics do not follow from the least action principle, although they partially retain some important features of Lagrangian systems. For example, covariant pre-symplectic structures have been originally proposed for Lagrangian field theories \cite{Crnkovic}, and they turn out to be a fruitful tool for studying their symmetries and conserved currents, especially for the gauge invariant models, see \cite{Havkine}. Pre-symplectic structures can be admitted not only by Lagrangian equations \cite{Sharapov-tricomplex}, which allows their useful properties to be applied beyond the class of variational dynamics, including, for example, higher-spin gravity models \cite{Alkalaev:2013hta}, \cite{Sharapov-presymplectic},  \cite{Dneprov:2024cvt}, \cite{Grigoriev:2021wgw}. Not necessarily Lagrangian classical equations of motion (EoM) can also admit another extra structure, termed the Lagrange anchor, which is dual, in a sense, to the pre-symplectic form. If the Lagrange anchor is admitted by the classical field equations, the dynamics can be equipped with classical BRST (Becchi-Rouet-Stora-Tyutin) complex which can be promoted to the quantum level either by deformation quantization \cite{Lyakhovich:2004xd}, or by covariant path integral methods \cite{Kazinski:2005eb}, \cite{Lyakhovich:2006sc}. For example, non-Lagrangian Donaldson-Uhlenbeck-Yau equations and anti-self-dual Yang-Mills theory can be covariantly quantized by these methods \cite{Lyakhovich:2007cw}. If the Lagrange anchor is known for non-Lagrangian equations, it connects symmetries and classical conserved quantities \cite{Kaparulin:2010ab}.

In this article, we study the classical field theory whose equations of motion follow from the least action principle, but the class of admissible trajectories is restricted by differential equations. Our method is explained in the next section, mostly by the case of the mechanical system with time $t$ being the only independent variable, and with a finite-dimensional configuration space. At the end of the next section, we explain how this extends to the field theory by reinterpreting the results in terms of De Witt's condensed notation.

Let us explain the problem setting.
We consider an under-determined system of $M$ ordinary differential equations (ODEs) restricting the trajectories $y^I(t),\, I=1, \ldots , N$ in the configuration space
\begin{equation}\label{eq_general}
T_A (y, \dot{y}, \ddot{y},  \cdots ) =0 \, , \qquad A=1, \ldots, M\,.
\end{equation}
By an under-determined system we mean a set of equations whose general solution contains arbitrary functions of $t$.

Consider an action functional
\begin{equation}\label{action_general}
    S[y(t)]=\int_{t_1}^{t_2} L(y, \dot{y}, \ddot{y},  \cdots) dt\,,
\end{equation}
where $L$ is a smooth function of $y, \dot{y}, \ddot{y},  \cdots, \overset{(r)}{y}{}^{I}$. The problem is to find the necessary conditions for the extrema of the functional (\ref{action_general}) in the class of trajectories satisfying equations
(\ref{eq_general}).

The Lagrange multiplier method is known to provide the necessary conditions for the conditional extrema \cite{Gelfand}. The method implies finding the unconditional extrema of the functional of the original variables $y(t)$ and Lagrange multipliers $\lambda (t)$
\begin{equation}\label{S-lambda}
 S_\lambda[y(t), \lambda(t)]=\int_{t_1}^{t_2} \left( L(y, \dot{y}, \ddot{y},
 \cdots) +\lambda^AT_A(y, \dot{y}, \ddot{y},  \cdots)\right) dt\, .
\end{equation}
Throughout the article, summation is assumed over repeated indices.
This action leads to the Euler-Lagrange equations for the extreme trajectories,
\begin{equation}\label{eq_lagrange_multiplier}
    \frac{\delta S_\lambda}{\delta y^I}=0\, ; \qquad \frac{\delta S_\lambda}{\delta \lambda^A}\equiv T_A(y, \dot{y}, \ddot{y},
 \cdots) =0 \, .
\end{equation}
In the next section, we suggest another solution to the problem of conditional extremum. It results in a system of equations for the extrema of any given functional of the form (\ref{action_general}) in the class of solutions to a general system (\ref{eq_general}). This system does not involve Lagrange multipliers or any other auxiliary variables, it is an ODE system solely for the original variables $y(t)$. In some special cases, the suggested system is equivalent to equations (\ref{eq_lagrange_multiplier}) in the sense that the solutions are the same for $y(t)$, and the number of independent Cauchy data coincide, even though the new system does not involve Lagrange multipliers. These special cases include the pure algebraic equations (\ref{eq_general}). One more case when our method is equivalent to the inclusion of Lagrange multipliers in the action is the class of equations (\ref{eq_general}) which describe a pure gauge system. This means that if the system (\ref{eq_general}) is considered as such, irrespectively to the action, all the degrees of freedom (DoFs) are gauged out by the gauge transformations of these equations. In general, the alternative equations for extrema of action (\ref{action_general}) restricted by (\ref{eq_general}) require less Cauchy data than equations (\ref{eq_lagrange_multiplier}).
Though the solutions for the original variables coincide, the Lagrange multipliers can require extra Cauchy data, in general. In this sense, the system with Lagrange multipliers has more DoFs than the suggested system, which we deduce in the next section.

The key element of the proposed construction is the complete gauge symmetry of the equations (\ref{eq_general}), as such, considered independently of an action. The unfree variation of the trajectories restricted by equations (\ref{eq_general})  reduces to the infinitesimal gauge symmetry transformation of these equations. We take the gauge variation of an action (\ref{action_general}) with respect to this symmetry. This gauge variation of an action necessarily vanishes on the extrema of the functional in the class of solutions of the differential equations (\ref{eq_general}). Given the gauge generators of equations (\ref{eq_general}), we explicitly derive the corresponding modification of the Lagrangian equations that follow from the requirement that this gauge variation of an action vanishes.

Suggested equations for extrema, being formulated without any recourse to Lagrange multipliers, admit Hamiltonian formulation (with degenerate Poisson bi-vector, in general), while the equations are not variational. The bracket is defined by the generators and structure functions of the gauge symmetry algebra of the differential equations, restricting the dynamics.

\section{Unfree variation of the action \texorpdfstring{\\ and the gauge symmetry of equations restricting the trajectories}{}}
By unfree variation $\tilde{\delta} y^I(t)$  we understand any infinitesimal transformation of the trajectory such that maps any solution of the equations (\ref{eq_general}) to another solution. This means that $\tilde{\delta} T_A$ vanishes on the trajectories such that obey equations (\ref{eq_general}),
\begin{equation}\label{delta-tilde-T}
  \tilde{\delta} T_A(y,\dot{y},\ddot{y}\, \dots)_{|T=0}=0\,.
\end{equation}
 The necessary condition for the extrema of a functional (\ref{action_general}) in the class of trajectories restricted by (\ref{eq_general}) is that any unfree variation of an action functional should vanish once equations (\ref{eq_general}) are satisfied \cite{Gelfand},
\begin{equation}\label{delta-tilde-S}
   \tilde{\delta} S_{|T=0}=0\, .
\end{equation}
Below, we propose a method for explicitly deriving equations for trajectories such that the above unfree variation of an action should necessarily vanish.
The idea of the method is to use the gauge symmetry of the equations (\ref{eq_general}) to explicitly construct the unfree variation in a parameterized form.

\subsection{Infinitesimal gauge symmetry of differential equations.}
Consider an infinitesimal $m$-parametric transformation of the trajectories $y^I$ with parameters $\epsilon^\alpha (t), \, \alpha=1, \ldots , m$ being arbitrary functions of time,
\begin{equation}\label{gauge_transformaion_general}
\delta_\epsilon y^I = \hat{R}^I_\alpha \epsilon^\alpha\, , \qquad \hat{R}^I_\alpha =\sum_{s=0}^k \stackrel{(s)}{R}{}^I_\alpha (y,\dot{y},\ddot{y}\, \cdots ) \frac{d^s}{dt^s},
\end{equation}
where $\stackrel{(s)}{R}{}^I_\alpha (y,\dot{y},\ddot{y}\, \cdots )$ are smooth functions.

The transformation (\ref{gauge_transformaion_general}) is considered as an infinitesimal gauge symmetry of equations (\ref{eq_general}) if it leaves them invariant. The gauge invariance means that
\begin{equation}\label{gauge_transformation_def}
   \delta_\epsilon T_A =\hat{U}{}_A^BT_B\approx 0\, , \quad \forall
   \epsilon^\alpha (t)\,,
\end{equation}
where $\hat{U}$ is a differential operator of finite order with coefficients depending on $y, \epsilon$ and their time derivatives. The symbol $\approx$ means the equality on-shell, i.e. modulo contributions that vanish on solutions (or - in other words - such that are proportional to $T$). The operators $\hat{R}{}_\alpha^I$ are termed gauge symmetry generators of equations (\ref{eq_general}) in the sense that any gauge transformation obeying (\ref{gauge_transformation_def})  is spanned by the action of operators (\ref{gauge_transformaion_general}) on some gauge parameters. The gauge generators are defined modulo natural equivalence relations. The generalities about gauge symmetries of variational equations can be found in \cite{Henneaux:1992ig}. For not necessarily variational equations, see
\cite{Lyakhovich:2004xd,Kazinski:2005eb}.

If equations (\ref{eq_general}) were variational, the Dirac-Bergmann algorithm \cite{Dirac} would provide finding all the infinitesimal gauge symmetries of the system. For general, not necessarily variational equations, the algorithm of finding the complete infinitesimal gauge symmetry is suggested in the article
\cite{Lyakhovich:2008hu}. With this regard, the infinitesimal gauge symmetry can be taken for known, given equations (\ref{eq_general}).

\subsection{Partially Lagrangian equations for conditional extrema of action.}
Once the gauge variation (\ref{gauge_transformaion_general}) satisfies relation (\ref{gauge_transformation_def}) with any parameters $\epsilon^\alpha(t)$, it maps any solution to a solution of the equations. The basic fact of the calculus of variations is that the variation of a functional must vanish on any extreme trajectory for any variation from the class of functions in which extrema are sought for \cite{Gelfand}.
This means, in particular, that the variation of functional (\ref{action_general}) under the gauge variation of $y(t)$ of the form (\ref{gauge_transformaion_general}) should vanish at the extreme point of the functional once $y(t)$ satisfy equations (\ref{eq_general}),
\begin{equation}\label{gauge-variation-S}
    \delta_\epsilon S = 0, \quad \forall\epsilon^\alpha (t); \quad
    T_A(y, \dot{y}, \ddot{y}, \cdots )=0 \, .
\end{equation}
 As one can see, infinitesimal gauge transformation (\ref{gauge_transformation_def}) has property (\ref{delta-tilde-T}) that defines the unfree variation of the trajectories, satisfying equations (\ref{eq_general}). Hence, the necessary condition for the extrema (\ref{delta-tilde-S}) requires the above gauge variation of action to vanish.

As far as gauge parameters $\epsilon (t)$ are arbitrary functions of time, this means the variational derivative of $S$ by $\epsilon$ should vanish at the critical point of an action in the class of trajectories subject to equations (\ref{eq_general}),
\begin{equation}\label{eq_conditional_extremum_general}
    \frac{\delta_\epsilon S}{\delta \epsilon^\alpha}\equiv
    \hat{R}^{\dagger I}_\alpha \frac{\delta S}{\delta y^I}=0 \, ,
\end{equation}
where $\frac{\delta S}{\delta y^I}$ is the Euler-Lagrange derivative in $y^I$ of functional (\ref{action_general}),
\begin{equation}
    \frac{\delta S}{\delta y^I}=\sum^{r}_{s=0}(-1)^s\frac{d^s}{dt^s}\left(\frac{\partial L}{\partial \overset{(s)}{y}{}^{I}}\right)\,,
\end{equation}
and $\hat{R}^{\dagger I}_\alpha$ is formal Hermitian conjugate operator to the gauge generator $\hat{R}^{I}_\alpha$ (\ref{gauge_transformaion_general}),
\begin{equation}\label{gauge_generator_conjugate}
\hat{R}^{\dagger I}_\alpha =\sum_{s=0}^k (-1)^s \frac{d^s}{dt^s}
\stackrel{(s)}{R}{}^I_\alpha (y,\dot{y},\ddot{y}\, \cdots ) \,.
\end{equation}
To summarize, the necessary condition for the extrema of functional (\ref{action_general}) under conditions (\ref{eq_general}) is to satisfy the system of equations (\ref{eq_general}), (\ref{eq_conditional_extremum_general}).
These equations are constructed making use of three main elements: (i) original differential equations (\ref{eq_general}) restricting the class of admissible trajectories; (ii) generators of infinitesimal gauge transformations for these equations (\ref{gauge_transformaion_general}), (\ref{gauge_transformation_def}); (iii) Lagrangian of an action functional (\ref{action_general}).

Let us make a couple of remarks on the equations we propose. Obviously, any solution of the usual Lagrangian equations obeys equations (\ref{eq_conditional_extremum_general}), not vice versa. The critical trajectories for the conditional extrema problem are not necessarily critical for an action with unrestricted trajectories, even if (\ref{eq_general}) are satisfied.

Equations (\ref{eq_lagrange_multiplier}) provide necessary conditions for the conditional extrema of a functional (\ref{action_general}) in terms of $y(t)$ and $\lambda (t)$ while equations (\ref{eq_general}), (\ref{eq_conditional_extremum_general}) (also defining critical trajectories) involve only original variables $y(t)$. Let us discuss the relations between these equations.

First, notice that equations (\ref{eq_general}), (\ref{eq_conditional_extremum_general}) are contained among equations (\ref{eq_lagrange_multiplier}). To see that, it is sufficient to contract equations $\frac{\delta S_\lambda}{\delta y}=0$ (\ref{eq_lagrange_multiplier}) with the gauge generator $R^\dagger$ (\ref{gauge_generator_conjugate}) and observe that $\lambda$'s drop out because of constraints (\ref{eq_general}) and relations (\ref{gauge_transformation_def}):
\begin{equation}\label{Rdagger-lambda}
\hat{R}^{\dagger I}_\alpha\frac{\delta S_\lambda}{\delta
y^I}=\hat{R}^{\dagger I}_\alpha \frac{\delta S}{\delta y^I} + \cdots,
\end{equation}
where $\cdots$ mean terms proportional to $T_A$. This demonstrates a uniform way of excluding the Lagrange multipliers and coming to a self-contained subsystem for extrema $y(t)$. 

Even though equations (\ref{eq_conditional_extremum_general}) are contained in the system with Lagrange multipliers (\ref{eq_lagrange_multiplier}), the latter system requires more Cauchy data in general. These two systems need the same number of initial conditions only in some special cases. These include a pure gauge system (\ref{eq_general}) and the case of algebraic constraints. In general, from the viewpoint of physics, the system with Lagrange multipliers (\ref{eq_lagrange_multiplier}) has more degrees of freedom than the system (\ref{eq_general}), (\ref{eq_conditional_extremum_general}). These extra degrees of freedom may have a substantial impact on dynamics. For example, the canonical energy for system (\ref{eq_lagrange_multiplier}) can be unbounded, while for equations (\ref{eq_general}), (\ref{eq_conditional_extremum_general}) it can be bounded. We will illustrate this feature in one of the examples below (\ref{dMdt_energy}), (\ref{dmdt_energy_lagrange_mult}).

Let us remark that the problem of conditional extremum can be considered in field theory as well as in mechanics. All the formulas are easily generalized to the case of field theory using condensed notation. In this notation, condensed labels include discrete indices and the space-time points, thus the field $\varphi^i(x)$, where $i=1\dots n$ can be written as $\varphi^I,\,I=(i,x)$. Summation over condensed labels includes summation over discrete indices and integration over spacetime variables. Gauge symmetry in field theory is written as (\ref{gauge_transformaion_general}), where $\hat{R}^I_\alpha$ is the kernel of the differential operator, and the time variable should be replaced with all spacetime coordinates $x^{\mu}$. The example of applying the algorithm of finding conditional extrema to the field theory is given in Section 4.

\section{Hamiltonian analysis}\label{Sec:3}
In the previous section, we have demonstrated that the necessary condition for the conditional extrema of an action functional in the class of trajectories restricted by differential equations (\ref{eq_general}) is that the trajectories are the solutions of the system of partially Lagrangian equations (\ref{eq_conditional_extremum_general}). In this section, we describe how to bring these equations to the Hamiltonian form.

Under certain regularity conditions, any system of ODEs by depressing the order  can be reduced to the following normal form:
\begin{equation}\label{normal_form_eq}
T^i\equiv    \dot{x}{}^i - Z^i_\alpha (x) u^\alpha-V^i=0 \, , \qquad
i=1, \ldots n\,, \quad \alpha = 1, \ldots  , m  \,.
\end{equation}
Variables $x^i$ can be considered as local coordinates on some manifold $\mathcal{M}$, $\{Z_\alpha\}$ is a basis in some vector distribution $\mathcal{Z}=\text{span}\{Z_\alpha\}$. $V^{i}$ are the components of vector field $\textbf{V}=V^{i}\partial_i$. This vector is termed drift. Given the normal form of equations (\ref{normal_form_eq}), the most general functional (\ref{action_general}) to extremize reads
\begin{equation}\label{L_x_u}
    S[x(t),u(t)]=\int_{t_1}^{t_2}L(x,u)dt\,,
\end{equation}
because $\dot{x}$ is defined by the equations. If $L$ involved derivatives of $u$, one could introduce further auxiliary variables absorbing the derivatives, and again bringing the equations to the normal form (\ref{normal_form_eq}). In \cite{Lyakhovich:2008hu}, the algorithm is provided for finding the complete set of gauge symmetry generators (\ref{gauge_transformaion_general}) for the equations in the normal form (\ref{normal_form_eq}). In principle, the order of the derivatives of gauge parameters in gauge variation (\ref{gauge_transformaion_general}), (\ref{gauge_transformation_def}) can be high enough \cite{Lyakhovich:2008hu} even though the equations (\ref{normal_form_eq}) are of the first order. This means the equations for conditional extrema (\ref{eq_conditional_extremum_general}) can be of a higher order. What is more, system (\ref{eq_conditional_extremum_general}), (\ref{normal_form_eq}) is non-variational in general. However, after depressing the order to the first, the system is Hamiltonian with drift. Let us detail the construction of the Hamiltonian form for equations (\ref{eq_conditional_extremum_general}), (\ref{normal_form_eq}). To provide the existence of unconstrained  Hamiltonian formalism, we impose the regularity requirement on $L$:
\begin{equation}\label{hessian}
    \det\left({\frac{\partial^2 L(x,u)}{\partial u^\alpha\partial
    u^\beta}}\right)\neq 0\,.
\end{equation}
The form of gauge symmetry of equations (\ref{normal_form_eq}) substantially depends on commutators between vector fields $Z_{\alpha}$ and $V$ in (\ref{normal_form_eq}).

At first, we consider the case of an integrable distribution $\mathcal{Z}$. A more general case is considered later. Integrable distribution means the following commutation relations between the generators:
\begin{equation}
\label{commutators_integrable}
    [Z_{\alpha},Z_{\beta}]=U^{\gamma}_{\alpha \beta} Z_{\gamma}\,, \quad [Z_{\alpha}, V]= V^{\beta}_{\alpha}Z_{\beta}\,,
\end{equation}
where $U^{\gamma}_{\alpha \beta}, V^{\beta}_{\alpha}$ are functions on manifold $\mathcal{M}$.
The basis vectors $Z_\alpha$ and the drift $V$ are supposed to be linearly independent, while the left-hand sides of the above relations are the commutators. Commuting these relations with $Z,V$ and applying the Jacobi identity, we arrive at the following relations between the structure functions,
\begin{equation}\label{UU}
\begin{gathered}
    U_{\alpha \beta}^{\gamma}V_{\gamma}^{\rho}+U_{\gamma\alpha}^{\rho}V_{\beta}^{\gamma}-U_{\gamma \beta}^{\rho}V_{\alpha}^{\gamma}-V^{j}\partial_j U_{\alpha \beta}^{\rho}-Z_{\alpha}^{j}\partial_j V_{\beta}^{\rho}+Z_{\beta}^{j}\partial_j V_{\alpha}^{\rho}=0\,,\\
    U_{\alpha \beta}^{\omega}U_{\omega \gamma}^{\rho}-Z^{j}_{\gamma}\partial_{j}U_{\alpha \beta}^{\rho}+\text{cycle}(\alpha, \beta, \gamma)=0\,.
\end{gathered}
\end{equation}
The gauge symmetry of equations (\ref{normal_form_eq}) in the case of integrable distribution (\ref{commutators_integrable}) reads \cite{Lyakhovich:2008hu}
\begin{equation}\label{gt-Ham-1}
    \delta_{\epsilon}x^i=Z^i_{\alpha} \epsilon^{\alpha}\,, \quad \delta_{\epsilon}u^{\alpha}=\dot{\epsilon}^{\alpha}-(V^{\alpha}_{\beta}+U^{\alpha}_{\beta \gamma}u^{\gamma})\epsilon^{\beta}\,.
\end{equation}
Given gauge transformations (\ref{gt-Ham-1}) of the equations (\ref{normal_form_eq}) for the case of integrable $\mathcal{Z}$, we apply relation (\ref{eq_conditional_extremum_general}) which leads to equations that are necessary conditions for trajectories to bring the action to the conditional extrema
\begin{equation}
\label{eq_conditional_extremum_involutive}
    \frac{\delta S}{\delta \epsilon^{\alpha}}=\frac{\partial L}{\partial x^{i}}Z^{i}_{\alpha}-\frac{d}{dt}\left(\frac{\partial L}{\partial u^{\alpha}}\right)-\frac{\partial L}{\partial u^{\beta}}(V^{\beta}_{\alpha}+U^{\beta}_{\alpha \gamma}u^{\gamma})\,.
\end{equation}

Let us introduce the momenta $p_\alpha$, being the Legendre transform of Lagrangian (\ref{L_x_u}) with respect to $u_\alpha$
and assume the non-degeneracy of Hessian (\ref{hessian}). In this case, variables $u^{\alpha}$ can be expressed in terms of phase space variables $x,p$,
\begin{equation}
\label{momentum_def}
    p_{\alpha}=\frac{\partial L(x,u)}{\partial u^{\alpha}} \iff u^{\alpha}=\overline{u}^{\alpha}(x,p)\,,
\end{equation}
where $L(x,u)=L(x,Z^{i}_{\alpha}u^{\alpha}+V^i)$.

Introduce the following Poisson brackets between the phase space variables:
\begin{equation}
    \{x^{i},x^{j}\}=0\,, \quad \{x^{i},p_{\alpha}\}=Z^{i}_{\alpha}\,, \quad \{p_{\alpha}, p_{\beta}\}=-U^{\gamma}_{\alpha \beta}\;p_{\gamma}\,.
\end{equation}
These brackets satisfy the Jacobi identity because of integrability conditions (\ref{commutators_integrable}), and their consequences (\ref{UU}).

Introduce the following Hamiltonian:
\begin{equation}
    \label{hamiltonian_gen}H(x,p)=p_{\alpha}\overline{u}^{\alpha}(x,p)-L(x,\overline{u}^{\alpha}(x,p))\,.
\end{equation}
Consider the equations following from this Hamiltonian with drift,
\begin{equation}
\begin{gathered}
    \dot{x}^i=\{x^{i},H\}+V^i\,,\\
    \dot{p}_{\alpha}=\{p_{\alpha},H\}-V^{\beta}_{\alpha}p_{\beta}\,.
\end{gathered}
\end{equation}
Using the definition of momenta (\ref{momentum_def}), one can see that the equations in the first line are equivalent to the system (\ref{normal_form_eq}). Equations in the second line are equivalent to equations (\ref{eq_conditional_extremum_involutive}). Then we showed that the formulation in terms of the Hamiltonian is equivalent to the Lagrangian formulation. Introduce the following vector field:
\begin{equation}
    \textbf{V}=V^{i}\frac{\partial}{\partial x^i}-V^{\beta}_{\alpha}p_{\beta} \frac{\partial}{\partial p_{\alpha}}\,.
\end{equation}
The Jacobi identity for the commutators of $Z$ and $V$ implies that this vector field differentiates the Poisson bracket, i.e
\begin{equation}
     \textbf{V}(\{O_1(x,p),O_2(x,p)\})=\{\textbf{V} O_1(x,p),O_2(x,p)\}+\{O_1(x,p),\textbf{V} O_2(x,p)\}\,,
\end{equation}
where $O_1$, and $O_2$ are the functions on the phase space. This property is quite important because the Poisson theorem still holds as in usual Hamiltonian mechanics: the Poisson bracket of two integrals of motion is the integral of motion.

Next, consider the case of non-integrable distribution $\{Z^{(0)}_{\alpha} \equiv Z_{\alpha} \}$ with the following commutation relations:
\begin{equation}
\begin{split}
\label{commutators_Z1}
  &[Z^{(0)}_{\alpha},Z^{(0)}_{\beta}]=U^{(00)\gamma}_{(0)\alpha \beta}Z^{(0)}_{\gamma} \,, \quad [Z^{(0)}_{\alpha},V]=Z^{(1)}_{\alpha}\,, \quad [Z^{(1)}_{\alpha},V]=V^{(1)\beta}_{(0)\alpha}Z^{(0)}_{\beta}+V^{(1)\beta}_{(1)\alpha}Z^{(1)}_{\beta}\,,\\
  &[Z^{(0)}_{\alpha},Z^{(1)}_{\beta}]=U^{(01)\gamma}_{(0)\alpha \beta}Z^{(0)}_{\gamma} +U^{(01)\gamma}_{(1)\alpha \beta}Z^{(1)}_{\gamma} \,, \quad [Z^{(1)}_{\alpha},Z^{(1)}_{\beta}]=U^{(11)\gamma}_{(0)\alpha \beta}Z^{(0)}_{\gamma} +U^{(11)\gamma}_{(1)\alpha \beta}Z^{(1)}_{\gamma} \,. \\
\end{split}
\end{equation}
This distribution becomes integrable if we include additional vector fields $\{Z^{(1)}_{\alpha}\}$, which are commutators between $Z$ and $V$, and the number of these vector fields coincides with the original ones. The gauge symmetry of equations (\ref{normal_form_eq}) in this case reads
\begin{equation}
\begin{split}
    &\delta_{\epsilon}x^{i}=\dot{\epsilon}^{\alpha}Z^{(0)i}_{\alpha}
    +\epsilon^{\alpha}\bigg(u^{\rho}Z^{(0)i}_{\omega}U^{(01)\omega}_{(1)\rho \alpha}+Z^{(1)i}_{\alpha}-Z^{(0)i}_{\omega}V^{(1)\omega}_{(1)\alpha}\bigg)\,,\\
    &\delta_{\epsilon}u^{\alpha}=\ddot{\epsilon}^{\alpha}+\dot{\epsilon}^{\beta}\bigg(u^{\rho}[U^{(01) \alpha}_{(1) \rho \beta}+U^{(00)\alpha}_{(0)\rho \beta}]-V^{(1)\alpha}_{(1)\beta}\bigg)+\epsilon^{\beta}\bigg(u^{\omega}u^{\rho}[U^{(00)\alpha}_{(0)\rho \gamma}U^{(01)\gamma}_{(1)\omega \beta}+Z^{(0)j}_{\rho}\partial_{j}U^{(01)\alpha}_{(1)\omega \beta}]\\
    &+u^{\rho}[U^{(01)\alpha}_{(0)\rho \beta}-U^{(00)\alpha}_{(0)\rho \gamma}V^{(1) \gamma}_{(1) \beta}-Z^{(0)j}_{\rho}\partial_{j}V^{(1)\alpha}_{(1)\beta}+V^{j}\partial_{j}U^{(01)\alpha}_{(1)\rho \beta}]+\dot{u}^{\rho}U^{(01)\alpha}_{(1)\rho \beta}-V^{(1)\alpha}_{(0)\beta}-V^{j}\partial_{j}V^{(1)\alpha}_{(1)\beta}\bigg)\,.
\end{split}
\end{equation}
Equations (\ref{eq_conditional_extremum_general}) in this case read as follows
\begin{equation}
\label{eq_conditional_extremum_Z1}
    \begin{split}
        &\frac{\delta S}{\delta \epsilon^{\alpha}}=\frac{d^2}{dt^2}\left(\frac{\partial L}{\partial u^{\alpha}}\right)-\frac{d}{dt}\left(\frac{\partial L}{\partial u^{\beta}}\right)\bigg(u^{\rho}[U^{(01)\beta}_{(1)\rho \alpha}+U^{(00)\beta}_{(0)\rho \alpha}]-V^{(1)\beta}_{(1)\alpha}\bigg)+\frac{\partial L}{\partial u^{\beta}}\bigg(u^{\omega}u^{\rho}[U^{(00)\beta}_{(0)\rho \gamma}U^{(01)\gamma}_{(1)\omega \alpha}\\
        &-Z^{(0)j}_{\omega}\partial_{j}U^{(00)\beta}_{(0)\rho \alpha}]-\dot{u}^{\rho}U^{(00)\beta}_{(0)\rho \alpha}+u^{\rho}[U^{(01)\beta}_{(0)\rho \alpha}-U^{(00)\beta}_{(0)\rho \gamma}V^{(1)\gamma}_{(1)\alpha}-V^{j}\partial_{j}U^{(00)\beta}_{(0)\rho \alpha}]-V^{(1)\beta}_{(0)\alpha}\bigg)\\
        &-\frac{d}{dt}\left(\frac{\partial L}{\partial x^{i}}Z^{(0)i}_{\alpha}\right)+\frac{\partial L}{\partial x^{i}}\bigg(u^{\rho}Z^{(0)i}_{\omega}U^{(01)\omega}_{(1)\rho \alpha}+Z^{(1)i}_{\alpha}-Z^{(0)i}_{\omega}V^{(1)\omega}_{(1)\alpha}\bigg)\,.
    \end{split}
\end{equation}
As in the case of the integrable distribution, define momenta (\ref{momentum_def})
\begin{equation}
    p^{(0)}_{\alpha}=\frac{\partial L(x,u)}{\partial u^{\alpha}} \iff u^{\alpha}=\overline{u}^{\alpha}(x,p^{(0)})\,,
\end{equation}
Introduce the Poisson brackets between the phase space variables:
\begin{equation}
\label{brackets_Z1}
    \begin{split}
        &\{x^{i},x^{j}\}=0\,, \quad \{x^{i},p^{(0)}_{\alpha}\}=Z^{(0)i}_{\alpha}\,, \quad \{x^{i},p^{(1)}_{\alpha}\}=Z^{(1)i}_{\alpha}\,,\quad \{p^{(0)}_{\alpha},p^{(0)}_{\beta}\}=-U^{(00)\gamma}_{(0)\alpha \beta}p^{(0)}_{\gamma}\,,\\
        &\{p^{(0)}_{\alpha},p^{(1)}_{\beta}\}=-U^{(01)\gamma}_{(0)\alpha \beta}p^{(0)}_{\gamma}-U^{(01)\gamma}_{(1)\alpha \beta}p^{(1)}_{\gamma}\,, \quad \{p^{(1)}_{\alpha},p^{(1)}_{\beta}\}=-U^{(11)\gamma}_{(0)\alpha \beta}p^{(0)}_{\gamma}-U^{(11)\gamma}_{(1)\alpha \beta}p^{(1)}_{\gamma}\,.
    \end{split}
\end{equation}
Here, in addition to original momenta $p^{(0)}$, we introduce extra momenta $p^{(1)}$ to bring the equations (\ref{eq_conditional_extremum_Z1}) to the first order form. Brackets (\ref{brackets_Z1}) satisfy the Jacobi identity as a consequence of commutation relations (\ref{commutators_Z1}).

The  Hamiltonian is introduced by the same rule as in the case of the integrable distribution $\mathcal{Z}$,
\begin{equation}
\label{hamiltonian_z1}
    H(x,p^{(0)})=p^{(0)}_{\alpha}\overline{u}^{\alpha}(x,p^{(0)})-L(x,\overline{u}(x,p^{(0)}))\,.
\end{equation}
The equations following from this Hamiltonian with drift read
\begin{equation}
\label{eq_hamiltonian_z1}
    \begin{split}
        &\dot{x}^{i}=\{x^{i},H\}+V^{i}\,,\\
        &\dot{p}^{(0)}_{\alpha}=\{p^{(0)}_{\alpha},H\}-p^{(1)}_{\alpha}\,,\\
        &\dot{p}^{(1)}_{\alpha}=\{p^{(1)}_{\alpha},H\}-V^{(1)\beta}_{(0)\alpha}p^{(0)}_{\beta}-V^{(1)\beta}_{(1)\alpha}p^{(1)}_{\beta}\,.
    \end{split}
\end{equation}
As in the integrable case, equations in the first line are equivalent to the original equations (\ref{normal_form_eq}), restricting the trajectories. The equations in the second line are just the definition of momenta $p^{(1)}$. These momenta absorb the higher derivatives of $x$ to bring the variational equations for conditional extrema (\ref{eq_conditional_extremum_Z1}) to the first-order form. The equations in the third line are equivalent to equations (\ref{eq_conditional_extremum_Z1}), given the definitions of momenta $p^{(0)}$ and $p^{(1)}$.

Consider the following vector field acting on the phase space with the variables $x, \, p^{(0)}, \,p^{(1)}$,
\begin{equation}
    \textbf{V}=V^{i}\frac{\partial}{\partial x^i}-p^{(1)}_{\alpha} \frac{\partial}{\partial p^{(0)}_{\alpha}}-(V^{(1)\beta}_{(0)\alpha}p^{(0)}_{\beta}+V^{(1)\beta}_{(1)\alpha}p^{(1)}_{\beta})\frac{\partial}{\partial p^{(1)}_{\alpha}}\,.
\end{equation}
This drift differentiates the Poisson brackets. This property is easy to see using the Jacobi identity that follows from the commutation relations (\ref{commutators_Z1}). It might be instructive to deduce this fact in one more way. Let us extend the phase space by introducing one more phase space variable $P$ with the following Poisson brackets:
\begin{equation}
    \{x^i,P\}=V^{i}\,, \quad \{p^{(0)}_{\alpha},P\}=-p^{(1)}_{\alpha}\,, \quad \{p^{(1)}_{\alpha},P\}=-V^{(1)\beta}_{(0)\alpha}p^{(0)}_{\beta}-V^{(1)\beta}_{(1)\alpha}p^{(1)}_{\beta}\,.
\end{equation}
The Jacobi identity for the additional brackets is satisfied as a consequence of the Jacobi identity for the commutators of vector fields $Z^{(0)},Z^{(1)}$ and the original drift $V$. Upon inclusion $P$, EoM (\ref{eq_hamiltonian_z1}) take the following form:
\begin{equation}
    \begin{split}
        &\dot{x}^{i}=\{x^{i},H+P\}\,,\\
        &\dot{p}^{(0)}_{\alpha}=\{p^{(0)}_{\alpha},H+P\}\,,\\
        &\dot{p}^{(1)}_{\alpha}=\{p^{(1)}_{\alpha},H+P\}\,.
    \end{split}
\end{equation}
This means that drift $\textbf{V}$ differentiates Poisson brackets (\ref{brackets_Z1}).

Now, let us briefly comment, without proof, on the case of general distribution $\mathcal{Z}$.  Consider the characteristic distribution $\mathcal{Z}$, which is not necessarily integrable. Denote the closure of the distribution by $\bar{\mathcal{Z}}$, $dim \bar{\mathcal{Z}} =\bar{m}> m$,
\begin{equation}\label{ZViter}
\bar{\mathcal{Z}} = \mathcal{Z}\cup [ \mathcal{Z}, \mathcal{Z}]\cup [\mathcal{Z},[\mathcal{Z},\mathcal{Z}]]\cdots\, .
\end{equation}
Let us further extend $\bar{\mathcal{Z}}$  by including all the commutators with the drift,
\begin{equation}\label{Z-bar}
  \bar{\mathcal{Z}}_V= \bar{\mathcal{Z}} \cup [\bar{\mathcal{Z}},V]\cup [[\bar{\mathcal{Z}},V],V]\cdots\,.
\end{equation}
Let us choose the basis $\{Z_{\bar{\alpha}}\}$ in $\bar{\mathcal{Z}}_V$ in such a way that the basis $\{Z_\alpha\}$ of the original characteristic distribution is included, $\{Z_{\bar{\alpha}}\}=\{Z_\alpha, Z_{\alpha'}\}$. By construction, $\bar{\mathcal{Z}}_V$ is an integrable distribution, so the commutation relations read
\begin{equation}
    [Z_{\bar{\alpha}},Z_{\bar{\beta}}]=U^{\bar{\gamma}}_{\bar{\alpha} \bar{\beta}}Z_{\bar{\gamma}}\,, \quad [Z_{\bar{\alpha}},V]=V^{\bar{\beta}}_{\bar{\alpha}}Z_{\bar{\beta}}\,.
\end{equation}
We impose the condition of non-degeneracy of Hessian (\ref{hessian}), introduce momenta (\ref{momentum_def}) and Hamiltonian (\ref{hamiltonian_gen}). We impose the Poisson brackets on the phase space variables,
\begin{equation}
\label{brackets_general}
    \{x^{i},x^{j}\}=0\,, \quad \{x^{i},p_{\bar{\alpha}}\}=Z^{i}_{\bar{\alpha}}\,, \quad \{p_{\bar{\alpha}},p_{\bar{\beta}}\}=-U^{\bar{\gamma}}_{\bar{\alpha} \bar{\beta}}p_{\bar{\gamma}}\,,
\end{equation}
where $\{p_{\bar{\alpha}}\}=\{p_{\alpha},p_{\alpha'}\}$. The EoM with drift take the form
\begin{equation}
\label{eq_hamiltonian_drift_general}
    \begin{split}
        \dot{x}^i&=\{x^{i}\,,H\}+V^{i}\,,\\
        \dot{p}_{\bar{\alpha}}&=\{p_{\bar{\alpha}}\,,H\}-V_{\bar{\alpha}}^{\bar{\beta}}p_{\bar{\beta}}\,.
    \end{split}
\end{equation}
The right-hand side of these equations includes the Poisson bracket with the Hamilton function and the drift vector field $\textbf{V}$, which is not necessarily Hamiltonian. This drift differentiates the Poisson bracket,
\begin{equation}\label{Drift-gen}
    \textbf{V}=V^i\frac{\partial}{\partial x^{i}}-V_{\bar{\alpha}}^{\bar{\beta}}p_{\bar{\beta}}\frac{\partial}{\partial p_{\bar{\alpha}}}\,.
\end{equation}
\begin{equation}\label{Drift-Leibnitz-gen}
\textbf{V}\{O_1(x,p),O_2(x,p)\}=\{\textbf{V}O_1(x,p),O_2(x,p)\}+ \{O_1(x,p), \textbf{V} O_2(x,p)\}\,.
\end{equation}
As we see, in terms of these variables, equations for extrema of the action (\ref{L_x_u}) in the class of trajectories restricted by equations (\ref{normal_form_eq}) are brought to the Hamiltonian form with drift. The drift is not necessarily a Hamiltonian vector field, but it differentiates the bracket by the Leibnitz rule. If the drift is the Hamiltonian vector field, the original equations of motion (\ref{eq_general}), (\ref{eq_conditional_extremum_general}) are brought to the Hamiltonian form, though they are not variational by construction. If the drift is not a Hamiltonian vector field, equations (\ref{eq_hamiltonian_drift_general}) still retain essential properties of Hamiltonian systems, due to (\ref{Drift-Leibnitz-gen}). In particular, conserved quantities still form a Poisson subalgebra. Given the drift that differentiates the bracket, the system with equations (\ref{eq_hamiltonian_drift_general}) admits deformation quantization \cite{Lyakhovich:2004xd}.

Let us consider the case of a pure gauge system, i.e. $dim \bar{\mathcal{Z}}=dim \mathcal{M}$ and $\bar{\mathcal{Z}}$ spans $T\mathcal{M}$, so $\det Z^i_{\bar{\alpha}} \neq 0$. Using the definition of Poisson brackets (\ref{brackets_general}) and Hamiltonian (\ref{hamiltonian_gen}),  equations (\ref{eq_hamiltonian_drift_general}) read
\begin{equation}
\label{eq_pure_gauge}
    \begin{split}
        &\dot{x}^{i}=Z^{i}_{\alpha}u^{\alpha}+V^{i}\,,\\
        &\dot{p}_{\bar{\alpha}}=Z^{i}_{\bar{\alpha}}\frac{\partial L}{\partial x^{i}} - U^{\bar{\gamma}}_{\bar{\alpha} \beta}p_{\bar{\gamma}}u^{\beta}-V^{\bar{\beta}}_{\bar{\alpha}}p_{\bar{\beta}}\,,\\
        &p_{\alpha}=\frac{\partial L}{\partial u^{\alpha}}\,.
    \end{split}
\end{equation}
We also included equations in the last line, which are just the definition of momenta (\ref{momentum_def}).
Let us introduce change of the original momenta $p_{\bar{\alpha}}$ by the new ones $\pi_{i}$ defined  by relations
\begin{equation}
    p_{\bar{\alpha}}=Z^{i}_{\bar{\alpha}}\pi_{i}\,.
\end{equation}
Here $\det Z^i_{\bar{\alpha}} \neq 0$, so this change of variables is invertible once in the pure gauge case $\dim\bar{\mathcal{Z}}=n$. In terms of variables $x^{i}, \pi_{j}$ Poisson brackets (\ref{brackets_general}) become canonical
\begin{equation}
    \{x^{i},x^{j}\}=0, \quad \{\pi_{i},\pi_{j}\}=0, \quad \{x^{i},\pi_{j}\}=\delta^{i}_{j}\,.
\end{equation}
EoM (\ref{eq_pure_gauge}) in terms of the canonical variables read
\begin{equation}
\label{eq_pontryagin}
\begin{split}
    &\dot{x}^{i}=Z^{i}_{\alpha}u^{\alpha}+V^{i}\,,\\
    &\dot{\pi}_{i}=\frac{\partial L}{\partial x^{i}} - \pi_{j}(\partial_i Z^{j}_{\alpha} u^{\alpha}+\partial_{i}V^j)\,,\\
    &Z^{i}_{\alpha}\pi_{i}=\frac{\partial L}{\partial u^{\alpha}}\,.
\end{split}
\end{equation}
These equations follow from the variational principle for the action
\begin{equation}\label{LMP}
    S[x(t),\pi(t), u(t)] =\int_{t_1}^{t_2}\bigg( \pi_i \left(\dot{x}{}^i- Z^i_\alpha (x) u^\alpha-V^{i}(x)\right) +L(x,u) \bigg) dt \, .
\end{equation}
Here $\pi_i$ serve as Lagrange multipliers for the equations in the first line of (\ref{eq_pontryagin}). Equations (\ref{eq_pontryagin}) are the Hamiltonian equations that follow from the Pontryagin maximum principle \cite{Agrachev}. For the case of a non-degenerate Poisson bi-vector, any differentiation of the bracket locally reduces to the Hamiltonian vector field. So the drift can be absorbed by the Hamilton function in this case.

To conclude the section, let us discuss distinctions between the Hamiltonian equations with drift (\ref{eq_hamiltonian_drift_general}) for conditional extrema of an action functional (\ref{L_x_u}) in the class of trajectories restricted by equations (\ref{normal_form_eq}), and equations (\ref{eq_pontryagin}) which follow from the requirement of unconditional extrema of an action (\ref{LMP}) with Lagrange multipliers $\pi_i$. If system (\ref{normal_form_eq}) is pure gauge (from the perspective of control theory, it means it is controllable),  equations (\ref{eq_hamiltonian_drift_general}) are equivalent to (\ref{eq_pontryagin}). In general, if $\dim\bar{\mathcal{Z}}=\bar{m}<n$, the equations (\ref{eq_hamiltonian_drift_general}) have $n+\bar{m}$ degrees of freedom, while equations with Lagrange multipliers (\ref{eq_pontryagin}) have $2n$. Obviously, these two systems are not equivalent unless $\bar{m}=n$.

\vspace{0.2 cm}
\textbf{A remark on the relation between the conditional extremum problem and D'Alembert's principle for Lagrangian systems subject to non-holonomic constraints}

It is well known that D'Alembert's principle, when applied to Lagrangian systems with non-holonomic constraints, does not, in general, solve the problem of finding a conditional extremum of the action~\cite{Goldstein}. We clarify this distinction by examining the underlying gauge symmetry associated with the constraint equations defining admissible trajectories.

The normal form~(\ref{normal_form_eq}) of first-order ODEs can always be equivalently recast in the Pfaffian form:
\begin{equation}\label{Pf}
T_a(x,\dot{x})\equiv  \theta_{ai}(x)\dot{x}^{i}-V_a(x)=0\, ,
\end{equation}
where the set of one-forms $\{\theta_a = \theta_{ai}(x) dx^i\}$ generates the annihilator of  the vector distribution $\mathcal{Z}$:
\begin{equation}\label{theta}
  \theta_{ai}(x)Z^i_\alpha (x) =0\, ,\quad V_a(x)=V^i(x)\theta_{ai}(x) \, .
\end{equation}
Conversely, given the Pfaffian system~(\ref{Pf}), one can always recover the normal form~(\ref{normal_form_eq}) using the kernel distribution $\mathcal{Z}$. In classical mechanics, non-holonomic constraints are typically formulated in Pfaffian form~\cite{Goldstein}. Instantaneous virtual displacements $\delta^\prime x^i$ are defined to lie in the kernel of the forms $\theta$:
\begin{equation}\label{delta-prime}
  \theta_{ai}(x)\delta^\prime x^i=0 \, .
\end{equation}
D'Alembert's principle then requires that
\begin{equation}\label{DA}
  \delta^\prime S[x(t)] =0\,  , \qquad S=\int L(x,\dot{x})\, dt
\end{equation}
for all variations satisfying~(\ref{delta-prime}). However, it is known~\cite{Goldstein} that such virtual displacements do not, in general, preserve the non-holonomic constraints~(\ref{Pf}). Thus, D'Alembert's principle addresses a different problem than the conditional extremum considered here. The partially Lagrangian equations~(\ref{eq_conditional_extremum_general}) represent necessary conditions for the conditional extremum of the action and differ fundamentally from those derived from D'Alembert's principle.

From the gauge symmetry perspective, the instantaneous virtual displacements may be parameterized by arbitrary time-dependent functions $\epsilon^\alpha(t)$:
\begin{equation}\label{deltaprimeZ}
  \delta^\prime x^i = Z_\alpha^i(x) \epsilon^\alpha \, .
\end{equation}
This provides a general solution to the constraint~(\ref{delta-prime}). The transformation~(\ref{deltaprimeZ}) constitutes a symmetry of the normal form equations~(\ref{normal_form_eq}) and, hence, of the equivalent non-holonomic constraints~(\ref{Pf}), provided that the vector distribution $\mathcal{Z}$ is integrable and preserves the drift vector field:
\begin{equation}\label{Z-int}
[\mathcal{Z},\mathcal{Z}]\subset\mathcal{Z}, \quad [V,\mathcal{Z}]\subset \mathcal{Z}. 
\end{equation}
In this special case, the gauge transformations coincide with virtual displacements, and the variational principle~(\ref{DA}) becomes equivalent to the conditional extremum problem.\footnote{The integrability conditions~(\ref{Z-int}) imply that the one-forms $\theta_{ai} dx^i$ generate a differential ideal $I$, and $dV_a \in I$. Consequently, there exist independent local functions $f_a(x)$ such that $Z^i_\alpha \partial_i f_a = 0$. Then the constraint equations~(\ref{Pf}) reduce to $\dot{f}_a - V_a(f) = 0$, meaning that the gauge-invariant quantities decouple from the dynamics at the kinematic level.}

In the non-integrable case, however, the gauge transformations of~(\ref{Pf}) involve time derivatives of the gauge parameters. Unlike the virtual displacements~(\ref{delta-prime}), the general gauge transformations~(\ref{gauge_transformaion_general}) preserve the constraints~(\ref{Pf}). In this broader setting, the partially Lagrangian equations~(\ref{eq_conditional_extremum_general}) are necessary for the conditional extremum, while D'Alembert's principle~(\ref{DA}) leads to different equations that are unrelated to the extremum problem.

It is also well known~\cite{Goldstein} that the variational principle~(\ref{DA}) can be reformulated by introducing Lagrange multipliers to enforce the conditions~(\ref{delta-prime}) that restrict variations to the kernel of $\theta_a$. This yields the equations:
\begin{equation}\label{D-L}
  \frac{\partial L(x,\dot{x})}{\partial x^i} - \frac{d}{dt} \frac{\partial L(x,\dot{x})}{\partial \dot{x}^i} - \lambda^a \theta_{ai}(x) = 0 \, .
\end{equation}
These differ from the equations~(\ref{eq_lagrange_multiplier}) that follow from requirement of the conditional extremum of the action, even if constraints~(\ref{Pf}) are the same. In the conditional extremum problem, the multipliers $\lambda^a$ are governed by a system of linear \textit{differential} equations with field-dependent coefficients~(\ref{eq_lagrange_multiplier}), whereas in~(\ref{D-L}), the multipliers satisfy a system of \textit{algebraic} equations.

The multipliers in~(\ref{D-L}) can be eliminated by contracting the equations with the null vectors $Z^i_\alpha$ of the rectangular matrix $\theta_{ai}$~(\ref{theta}), leading to a closed system of second-order equations:
\begin{equation}\label{D-Z}
Z^i_\alpha(x) \left( \frac{\partial L(x,\dot{x})}{\partial x^i} - \frac{d}{dt} \frac{\partial L(x,\dot{x})}{\partial \dot{x}^i} \right) = 0 \, .
\end{equation}
These equations are, in general, different from the partially Lagrangian equations~(\ref{eq_conditional_extremum_general}), even when both the constraints~(\ref{Pf}) and the action are the same. They coincide only when the non-holonomic constraints~(\ref{Pf}) are integrable as in~(\ref{Z-int}). In that special case, the constraints describe a subsystem that decouples kinematically from the rest of the dynamics governed by the second-order equations~(\ref{D-Z}). In the non-integrable case, this kinematic decoupling fails because the gauge transformations involve time derivatives of the parameters.

To eliminate the multipliers in the conditional extremum equations~(\ref{eq_lagrange_multiplier}), one must act on their left-hand sides with differential operators $\hat{R}^\dagger$~(\ref{Rdagger-lambda}), which are the formal Hermitian conjugates of the gauge symmetry generators~(\ref{gauge_transformaion_general}). Importantly, this elimination procedure does not require the constraints to be written in any special form — such as the Pfaffian form~(\ref{Pf}) — but only requires knowledge of the gauge symmetry of the equations defining admissible trajectories. This makes the method particularly useful in field theory, where one seeks to derive field equations~(\ref{eq_conditional_extremum_general}) in a manifestly covariant fashion.

In summary, while D'Alembert's principle is a useful tool in classical mechanics, it addresses a problem distinct from that of the conditional extremum of the action functional.

\section{Examples}\label{Sec:4}
In this section, we illustrate the general formalism developed in the article by two examples of the systems where the extrema of the commonly known simple actions are sought for in the class of trajectories restricted by reasonable equations.
\subsection{Particle in the central field with trajectories restricted by conservation of angular momentum.}
Consider the plane motion of a non-relativistic point particle in a central field. The action reads
\begin{equation}
\label{dMdt_action}
    S[r(t),\varphi(t)]=\int_{t_1}^{t_2}  \left(\frac{m \dot{r}^2}{2}+\frac{mr^2 {\dot{\varphi}}^2}{2}-U(r)\right)dt\,,
\end{equation}
where $r$ is the radial coordinate, $\varphi$ is the polar angle and $U(r)$ is the potential energy. Let us find the extrema of this action under the following condition
\begin{equation}
\label{dMdt}
    \frac{dM}{dt}=0\,, \quad M=m r^2 \dot{\varphi}\,.
\end{equation}
This is the condition of conservation of angular momentum. This equation is part of the usual Lagrange equations for the action (\ref{dMdt_action}). This does not mean, the conditional extrema are the same as the unconditional critical trajectories. If the class of varying trajectories is restricted by this equation, there can be some extra critical trajectories, as we see below.

Gauge symmetry transformation of eq. (\ref{dMdt}), being considered irrespectively to the action reads
\begin{equation}
\label{dMdt_gauge_sym}
    \delta_{\epsilon} r=\dot{r}{\epsilon}-\frac{r}{2}\dot{\epsilon}\,,\quad \delta_{\epsilon}\varphi=\dot{\varphi}\epsilon\,.
\end{equation}
Variational derivatives of the action (\ref{dMdt_action}):
\begin{equation}
    \frac{\delta S}{\delta r}= -m\ddot{r}+mr{\dot{\varphi}}^2-\frac{dU}{dr}\,, \quad \frac{\delta S}{\delta \varphi}=-\frac{dM}{dt} \,.
\end{equation}
Substituting these variational derivatives and gauge symmetry (\ref{dMdt_gauge_sym})
into the general equations for conditional extrema (\ref{eq_conditional_extremum_general}), we arrive at the equation
\begin{equation}\label{dMdt_condtitional_equation}
    \frac{d}{dt}\left(\frac{m\Ddot{r}r}{2}+\frac{m{\dot{r}}^2}{2}+U(r)+\frac{r}{2}\frac{d U}{d r}\right)=0\, .
\end{equation}
This equation is of the third order, so it requires one more initial condition compared to the Lagrangian equations for unconditional extrema of the action for the particle in central potential (\ref{dMdt_action}). This is because equation (\ref{dMdt}), being a condition imposed to select admissible trajectories, admits solutions with any acceleration, not necessarily obeying the usual Lagrangian equations. Being a total derivative, the equation (\ref{dMdt_condtitional_equation}) can be immediately integrated
\begin{equation}\label{E}
\frac{m\Ddot{r}r}{2}+\frac{m{\dot{r}}^2}{2}+U(r)+\frac{r}{2}\frac{d U}{d r}=E\,,
\end{equation}
where $E$ is the constant of integration.

Let us introduce a new variable $x=\frac{r^2}{4}$. For this variable, relation (\ref{E}) is just a Newtonian equation for one-dimensional motion,
\begin{equation}
\label{dMdt_conditional_newton_form}
    m\Ddot{x}=-\frac{d\widetilde{U}(x)}{d x}\,,
\end{equation}
where
\begin{equation}
    \widetilde{U}(x)=x[U(r(x))-E]\,.
\end{equation}
The general solution to eq. (\ref{dMdt_conditional_newton_form}) reads
\begin{equation}
    \widetilde{E}=\frac{m{\dot{x}}^2}{2}+\widetilde{U}(x)=\frac{mr^2{\dot{r}}^2}{8}+\frac{U(r)r^2}{4}-\frac{Er^2}{4}\,.
\end{equation}
Choosing the arbitrary constant of integration in the form $\widetilde{E}=K-\frac{M^2}{8m}$ and using the equation (\ref{dMdt_condtitional_equation}), we find new conserved quantity
\begin{equation}\label{K}
    K=\frac{r^3}{8}(-m\ddot{r}-\frac{dU}{dr}+\frac{M^2}{mr^3})\,.
\end{equation}
We term this extra conserved quantity \emph{precession parameter}.
The value of  $K$ determines the acceleration of the particle at the initial time moment. Since the equations of motion (\ref{dMdt_condtitional_equation}) are of the third order, this quantity can be chosen arbitrarily for different solutions. Solutions with $K=0$ are admissible, and they correspond to the usual solutions for a particle in a central potential. With this definition of precession parameter $K$, the integral of motion $E$ takes the following form
\begin{equation}\label{dMdt_energy}
    E=\frac{m{\dot{r}}^2}{2}+\frac{M^2-8Km}{2mr^2}+U(r)\,.
\end{equation}
It is the usual expression for the energy of a non-relativistic particle in the central field modified with the precession parameter $K$. The term
\begin{equation}\label{centrifuga}
    \frac{M^2-8Km}{2mr^2}
\end{equation}
can be viewed as a centrifugal energy of the usual motion of the particle in a central field with the unrestricted class of trajectories, and with the constant in the numerator modified by the precession parameter $K$. If the numerator remains positive, i.e. $K<\dfrac{M^2}{8m}$, and the potential is not rapidly decreasing at $r\mapsto 0$, the energy (\ref{dMdt_energy}) is bounded from below in the vicinity of the center. Hence, the particle does not fall into the center, much like the usual case of the central field, though the trajectories can change because of the shifted constant in the centrifugal energy.

Consider the particular example of central field - Coulomb potential $U(r)=-\frac{\alpha}{r}, \; \alpha>0$. Integrating equations (\ref{dMdt}), (\ref{dMdt_energy}) we derive the trajectory $r(\varphi)$. There can be different types of trajectory, depending on the value of the precession parameter and energy,
\begin{equation}
\label{dMdt_trajectory}
\begin{gathered}
r(\varphi)=
    \begin{cases}
			\dfrac{p}{1+e\cos{(\frac{\varphi-\varphi_0}{\gamma})}}\,, & \text{if }K< \dfrac{M^2}{8m}\\[5 mm]
            \dfrac{2M^2 \alpha}{m\alpha ^2 (\varphi-\varphi_0) ^2 - 2 M^2E}, & \text{if } K = \dfrac{M^2}{8m}
		 \end{cases}
\end{gathered}\,.
\end{equation}
\begin{equation}
\label{dMdt_trajectory2}
    \begin{gathered}
        r(\varphi)=
    \begin{cases}
			\dfrac{p}{e^2\cosh{(\frac{\varphi-\varphi_0}{\gamma})}-e\sqrt{e^2-1}\sinh{(\frac{\varphi-\varphi_0}{\gamma})}-1}, & \text{if } E<0\\[5 mm]
            \dfrac{p}{e\cosh{(\frac{\varphi-\varphi_0}{\gamma})}-1}, & \text{if } 0\le E<\dfrac{m\alpha^2\gamma^2}{2M^2}\\[5mm]
            \dfrac{p \,\mathrm{e}^{\frac{\varphi-\varphi_0}{\gamma}}}{1-\mathrm{e}^{\frac{\varphi-\varphi_0}{\gamma}}}, & \text{if } E=\dfrac{m\alpha^2\gamma^2}{2M^2}\\[5mm]
            \dfrac{p}{e\sinh{(\frac{\varphi_0-\varphi}{\gamma})}-1}\,,& \text{if } E>\dfrac{m\alpha^2\gamma^2}{2M^2}
		 \end{cases}
    \end{gathered}\,,K> \dfrac{M^2}{8m}\,,
\end{equation}
where
\begin{equation}
\gamma = \frac{1}{\sqrt{\left|1-\frac{8Km}{M^2}\right|}}\,, \quad e=\sqrt{\left|1+\text{sign}{\left(1-\frac{8Km}{M^2}\right)}\frac{2EM^2}{\gamma^2 m \alpha^2}\right|}\,, \quad p=\frac{M^2}{\gamma^2 m \alpha}
\end{equation}
 and $\varphi_0$ is the angle at the initial time. In the case $K<\dfrac{M^2}{8m}$, the trajectory of the particle is a conic section with precession, the value of which depends on the precession parameter $K$. When $K\ge \dfrac{M^2}{8m}$, the particle falls into the center or goes to infinity (depending on initial conditions) along a spiral-like trajectory. Two examples of trajectory (\ref{dMdt_trajectory}), (\ref{dMdt_trajectory2}) can be seen at the figures:
\begin{figure}[H]
   \begin{minipage}{0.48\textwidth}
     \centering
     \includegraphics[width=.7\linewidth]{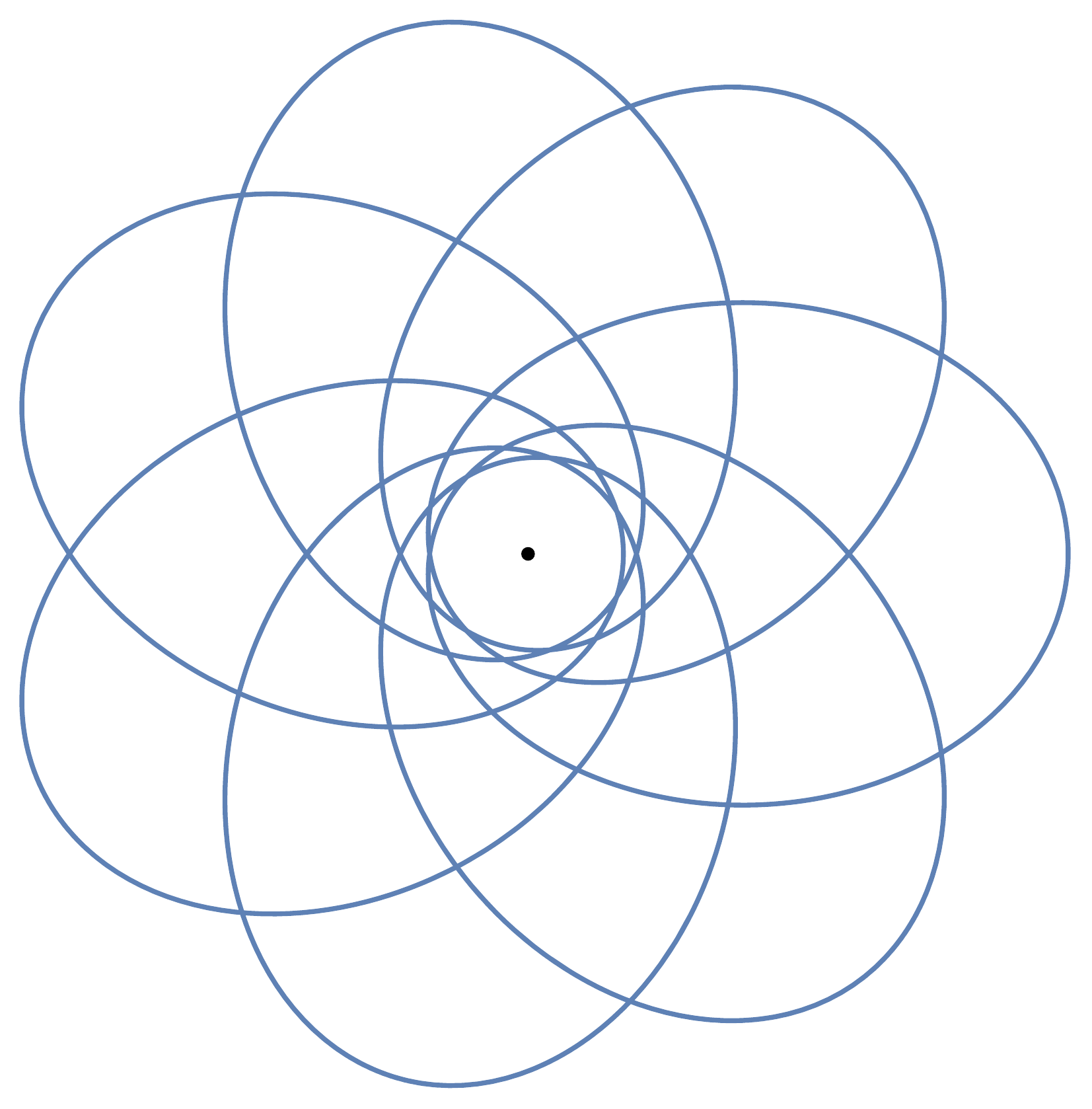}
     \caption{Precessing elliptical trajectory with parameters \\$K<\frac{M^2}{8m}\,,e=0.7\,,\gamma=8/7\,.$}\label{Fig:Data1}
   \end{minipage}\hfill
   \begin{minipage}{0.48\textwidth}
     \centering
     \includegraphics[width=.7\linewidth]{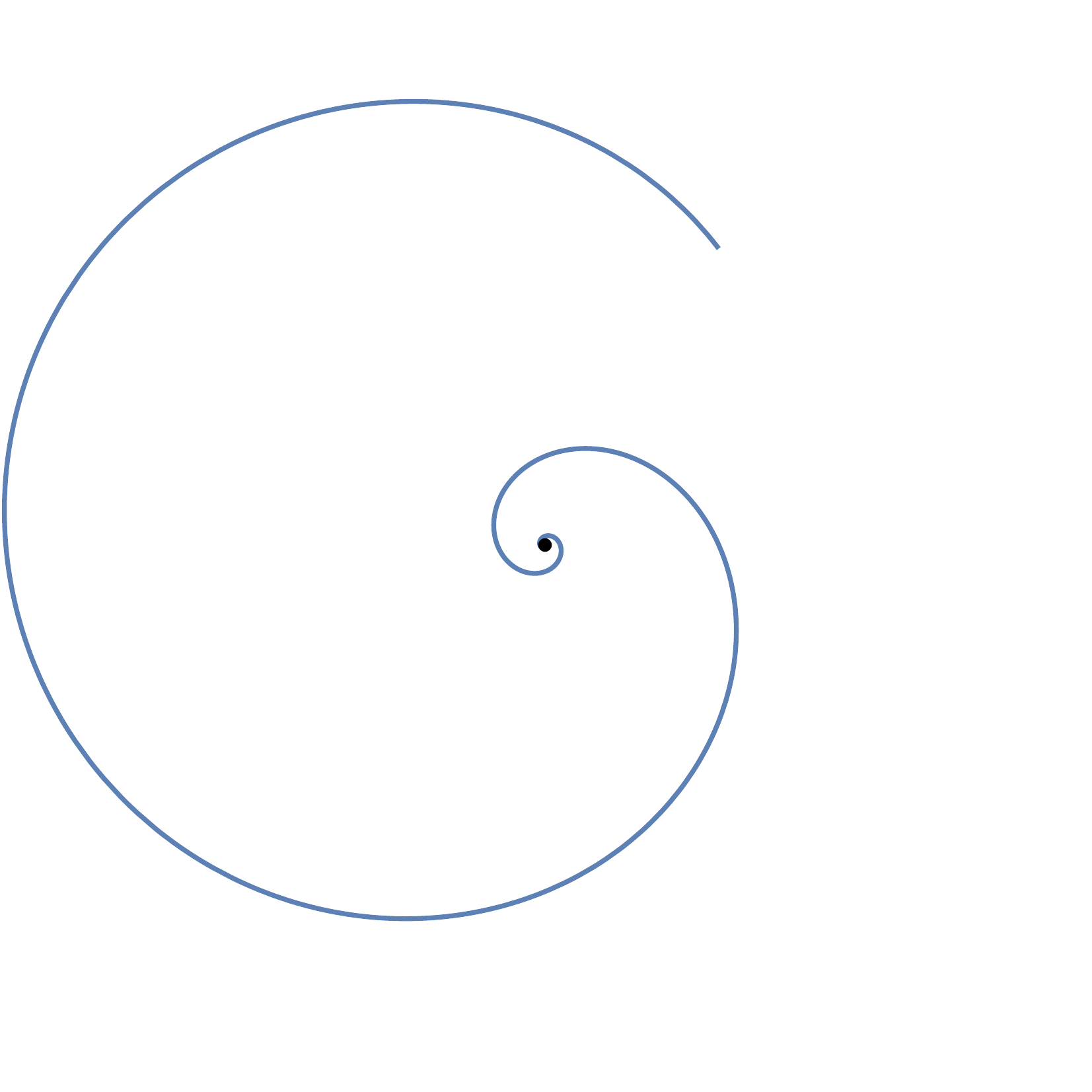}
     \caption{The spiral-like trajectory of the particle falling to the center with parameters $K>\frac{M^2}{8m}\,, E<0\,,e=1.5\,,\gamma=3\,.$}\label{Fig:Data2}
   \end{minipage}
\end{figure}
The problem of extrema for the action of a particle in a central potential  (\ref{dMdt_action})  in the class of trajectories restricted by the condition of conserving angular momentum (\ref{dMdt}) should admit a Hamiltonian formulation. Let us construct the Hamiltonian formalism following the general recipe of the previous section. Denote $x^1=r, x^2=\varphi, x^3=M$.
Equation (\ref{dMdt}) can be brought to the first-order normal form (\ref{normal_form_eq}),
\begin{equation}
\label{dMdt_normal}
    \dot{x}^{i}-Z^{i}u-V^i=0\,, \quad Z=
    \begin{pmatrix}
1 \\
0 \\
0
\end{pmatrix}  \,,\quad
V= \begin{pmatrix}
0 \\
\frac{M}{mr^2} \\
0
\end{pmatrix}  \,.
\end{equation}
The distribution formed by vector fields $Z,V$ is not integrable. Closure of this distribution contains one more vector field $Z_1$,
\begin{equation}
       [Z,V]=Z_1\,, \quad [Z_1,V]=0\,, \quad  [Z,Z_1]=-\frac{3}{r}Z_1\,, \quad  Z_1= \begin{pmatrix}
0 \\
-\frac{2M}{mr^3} \\
0
\end{pmatrix}  \,.
\end{equation}
Gauge symmetry of equations (\ref{dMdt_normal}) read
\begin{equation}
        \delta_{\epsilon}r=\dot{\epsilon}-\frac{3}{r}u\epsilon\,, \quad \delta_{\epsilon}\varphi=-\frac{2M}{mr^3}\epsilon\,, \quad \delta_{\epsilon}M=0\,, \quad
        \delta_{\epsilon}u=\ddot{\epsilon}-\frac{3}{r}u\dot{\epsilon}+\frac{3}{r}\left(\frac{u^2}{r}-\dot{u}\right)\epsilon\,.
\end{equation}
Following the general recipe of the previous section (\ref{brackets_Z1}),
we introduce the Poisson brackets between phase space variables $r, \varphi, M, p, p_1$
\begin{equation}
    \{r,p\}=1\,,\quad \{\varphi,p_1\}=-\frac{2M}{mr^3}\,,\quad \{p,p_1\}=\frac{3}{r}p_1\,.
\end{equation}
The  Hamiltonian is constructed by the rule (\ref{hamiltonian_z1})
\begin{equation}
    H(x,p)=\frac{p^2}{2m}-\frac{M^2}{2mr^2}+U(r)\,.
\end{equation}
The EoM are
\begin{equation}
    \begin{split}
        &\dot{r}=\{r,H\}=\frac{p}{m}\,,\\
        &\dot{\varphi}=\{\varphi,H\}+\frac{M}{mr^2}=\frac{M}{mr^2}\,,\\
        &\dot{M}=\{M,H\}=0\,,\\
        &\dot{p}=\{p,H\}-p_1=-\frac{M^2}{mr^3}-\frac{dU}{dr}-p_1\,,\\
        &\dot{p_1}=\{p_1,H\}=-\frac{3pp_1}{mr}\,.
    \end{split}
\end{equation}
These EoM include drift (\ref{dMdt_normal}). In fact, this drift is the Hamiltonian vector field. To demonstrate that, let us introduce a new Hamiltonian
\begin{equation}\label{H-prime}
    H^{\prime}=H-\frac{r}{2}p_1=\frac{p^2}{2m}-\frac{M^2}{2mr^2}-\frac{r}{2}p_1+U(r)\,.
\end{equation}
This Hamiltonian, being restricted on the level surface of conserved precession parameter $K$, is just energy (\ref{dMdt_energy}) expressed in terms of the phase space variables
\begin{equation}
    H^{'}\big|_{K=\text{const}}=\frac{p^2}{2m}+\frac{M^2-8Km}{2mr^2}+U(r)\,.
\end{equation}
It is bounded from below if the potential is not rapidly decreasing at $r\mapsto 0$ and $K<\dfrac{M^2}{8m}$ as we discussed after formula (\ref{dMdt_energy}).

With Hamilton function (\ref{H-prime}), the phase space equations of motion become Hamiltonian,
\begin{equation}
    \dot{x}^i=\{x^i,H^{\prime}\}\,, \quad \dot{p}=\{p,H^{\prime}\}\,, \quad \dot{p_1}=\{p_1,H^{\prime}\}\,.
\end{equation}
There are three integrals of motion in involution: energy, angular momentum, and precession parameter,
\begin{equation}
    \begin{split}
        &M\,, \quad H^{'}=E\,, \quad K=\frac{M^2}{4m}+\frac{r^3}{8}p_1; \quad \{M,H^{'}\}=\{K,H^{'}\}=\{M,K\}=0\,.
    \end{split}
\end{equation}
 Consider now the conditional extrema of the action (\ref{dMdt_action}) by introducing the Lagrange multiplier to the condition (\ref{dMdt}),
 \begin{equation}\label{dMdtS-lambda}
     S_{\lambda}[r(t),\varphi(t),\lambda(t)] =\int  \left(\frac{m \dot{r}^2}{2}+\frac{mr^2 {\dot{\varphi}}^2}{2}-U(r)-\dot{\lambda}mr^2\dot{\varphi}\right)dt\,.
 \end{equation}
 For the action with multipliers, the equations read
 \begin{equation}\label{eq-ex1-lambda}
     \frac{\delta S_{\lambda}}{\delta r}= -m\ddot{r}+mr{\dot{\varphi}}^2-\frac{dU}{dr}-2mr\dot{\lambda}\dot{\varphi}=0\,, \quad \frac{\delta S_{\lambda}}{\delta \varphi}=-\frac{d}{dt}(M-mr^2\dot{\lambda})=0 \,, \quad \frac{\delta S_{\lambda}}{\delta \lambda}=\frac{dM}{dt}=0\,.
 \end{equation}
 Obviously, there are three second-order equations, so the system implies six Cauchy data. The system without Lagrange multipliers includes the second-order equation (\ref{dMdt}) and one more equation (\ref{dMdt_condtitional_equation}) of the third order,  so five initial conditions are required. The extra degree of freedom is brought into the system by the Lagrange multiplier. It is unrelated to the motion of this particle as such.

Consider the canonical energy for the action with Lagrange multipliers (\ref{dMdtS-lambda})
\begin{equation}
\label{dmdt_energy_lagrange_mult}
    E=\frac{\partial L}{\partial \dot{r}}\dot{r}+\frac{\partial L}{\partial \dot{\varphi}}\dot{\varphi} + \frac{\partial L}{\partial \dot{\lambda}}\dot{\lambda}-L=\frac{m\dot{r}^2}{2}+\frac{mr^2(\dot{\varphi}-\dot{\lambda})^2}{2}-\frac{mr^2\dot{\lambda}^2}{2}+U(r)\,.
\end{equation}
The energy is unbounded from below since there is a negative kinetic term due to the Lagrange multiplier.
Evolution of $r$ can be bounded, or unbounded, depending on the potential energy and the initial data for the precession parameter $K$ (see (\ref{K})), while the above energy is always unbounded due to the contribution of the non-physical mode brought into the system by the Lagrange multiplier.

\subsection{Linearized gravity and Nordstr\"{o}m equation.}

Consider the linearized Einstein-Hilbert action
\begin{equation}
\label{EH_linearized}
    S_{\text{EH}}=\int d^4x\,\mathcal{L}_{\text{EH}}=\frac{1}{4}\int d^4x(\partial_{\mu}h_{\nu \lambda}\partial^{\mu}h^{\nu \lambda}+2\partial^{\mu}h\partial^{\nu}h_{\nu \mu}-2\partial^{\mu}h^{\nu \lambda}\partial_{\lambda}h_{\nu \mu}-\partial_{\mu}h\partial^{\mu}h+4\Lambda h)\,,
\end{equation}
where $\eta_{\mu \nu}$ is the Minkowski metric, $h_{\mu \nu}$ is the small perturbation of Minkowski metric, $h=\eta^{\mu\nu}h_{\mu\nu}$, and $\Lambda$ is the cosmological constant.
The linearized Einstein equations read
\begin{equation}
    \frac{\delta S_{\text{EH}}}{\delta h^{\mu \nu}} \equiv \frac{1}{2}(\partial_{\mu}\partial^{\lambda}h_{\nu \lambda}+\partial_{\nu}\partial^{\lambda}h_{\mu \lambda}- \Box h_{\mu \nu} -\partial_{\mu}\partial_{\nu}h)-\frac{1}{2}\eta_{\mu \nu}(\partial^{\lambda}\partial^{\rho}h_{\lambda \rho}-\Box h)+\Lambda \eta_{\mu \nu}=0\,.
\end{equation}
They have an algebraic consequence -- linearized Nordstr\"{o}m equation
\begin{equation}
\label{nordstrom_linearized}
   N\equiv -\eta^{\mu \nu}\frac{\delta S_{\text{EH}}}{\delta h^{\mu \nu}} = \partial^{\alpha}\partial^{\beta}h_{\alpha \beta}- \Box h-4\Lambda=0\,.
\end{equation}
 Its gauge symmetry read \cite{Abakumova:2022eoo}
\begin{equation}\label{norstrom-gauge-sym}
    \delta h^{\mu \nu}=\partial_{\lambda}H^{\mu \nu \lambda}-\frac{1}{3}\eta_{\alpha \beta}(\eta^{\mu \nu}\partial_{\lambda}H^{\alpha \beta \lambda}+\partial^{\nu}H^{\alpha \beta \mu}+\partial^{\mu}H^{\alpha \beta \nu})\,,
\end{equation}
where $H^{\mu \nu \lambda}$ is the gauge parameter with the hook symmetry in the symmetric basis,
\begin{equation}
    H^{(\mu \nu)\lambda}=H^{\mu \nu \lambda},\quad H^{(\mu \nu \lambda)}=0\,,
\end{equation}
where round brackets mean symmetrization of corresponding indices.
We rewrite this gauge transformation with explicit hook symmetry,
\begin{equation}
    \delta h^{\mu \nu}=\rho^{\mu \nu}_{\alpha \beta \gamma}H^{\alpha \beta \gamma}\,,
\end{equation}
where
\begin{equation}
\begin{split}
    \rho^{\mu \nu}_{\alpha \beta \gamma}=&\frac{1}{3}\Bigl[\delta^{\mu}_{\alpha}\delta^{\nu}_{\beta}\partial_{\gamma}+\delta^{\mu}_{\beta}\delta^{\nu}_{\alpha}\partial_{\gamma}-\delta^{\mu}_{\beta}\delta^{\nu}_{\gamma}\partial_{\alpha}-\delta^{\mu}_{\gamma}\delta^{\nu}_{\beta}\partial_{\alpha}-\frac{2}{3}\eta_{\alpha \beta}(\eta^{\mu \nu}\partial_{\gamma}+\delta^{\mu}_{\gamma}\partial^{\nu}+\delta^{\nu}_{\gamma}\partial^{\mu})\\
    &+\frac{2}{3}\eta_{\beta \gamma}(\eta^{\mu \nu}\partial_{\alpha}+\delta^{\mu}_{\alpha}\partial^{\nu}+\delta^{\nu}_{\alpha}\partial^{\mu})\Bigr]\,.
\end{split}
\end{equation}
The problem is to find the necessary conditions for the extrema of the action (\ref{EH_linearized}) in the class of functions satisfying equation (\ref{nordstrom_linearized}). The EoM for conditional extrema take the following form
\begin{equation}
\begin{split}
\label{eq_nordstrom_conditional}
        L_{\alpha \beta \gamma} \equiv -{\rho}{}^{\mu \nu}_{\alpha \beta \gamma} \frac{\delta S_{\text{EH}}}{\delta h^{\mu \nu}}&=-\frac{1}{3}\Bigl[\delta^{\mu}_{\alpha}\delta^{\nu}_{\beta}\partial_{\gamma}+\delta^{\mu}_{\beta}\delta^{\nu}_{\alpha}\partial_{\gamma}-\delta^{\mu}_{\beta}\delta^{\nu}_{\gamma}\partial_{\alpha}-\delta^{\mu}_{\gamma}\delta^{\nu}_{\beta}\partial_{\alpha}-\frac{2}{3}\eta_{\alpha \beta}(\eta^{\mu \nu}\partial_{\gamma}+\delta^{\mu}_{\gamma}\partial^{\nu}+\\
        &\delta^{\nu}_{\gamma}\partial^{\mu})+\frac{2}{3}\eta_{\beta \gamma}(\eta^{\mu \nu}\partial_{\alpha}+\delta^{\mu}_{\alpha}\partial^{\nu}+\delta^{\nu}_{\alpha}\partial^{\mu})\Bigr]\frac{\delta S_{\text{EH}}}{\delta h^{\mu \nu}}=-\frac{2}{3}\Bigl[\partial_{\gamma}\frac{\delta S_{\text{EH}}}{\delta h^{\alpha \beta}}-\partial_{\alpha}\frac{\delta S_{\text{EH}}}{\delta h^{\beta \gamma}}\\
        &-\frac{1}{3}\eta^{\mu \nu}(\eta_{\alpha \beta}\partial_{\gamma}\frac{\delta S_{\text{EH}}}{\delta h^{\mu \nu}}-\eta_{\beta \gamma} \partial_{\alpha}\frac{\delta S_{\text{EH}}}{\delta h^{\mu \nu}})\Bigr]=0\,,\\
        N \equiv -\eta^{\mu \nu}\frac{\delta S_{\text{EH}}}{\delta h^{\mu \nu}} &= \partial^{\alpha}\partial^{\beta}h_{\alpha \beta}- \Box h-4\Lambda=0\,.
\end{split}
\end{equation}
All the Einstein solutions, i.e. such metrics that  $\frac{\delta S_{\text{EH}}}{\delta h^{\mu \nu}}=0$, obey this system. However, there could be more solutions, as the above equations are of the third order, so there is a question about the degree of freedom for this system. In this work, we use the method of counting degrees of freedom proposed in \cite{Kaparulin:2012px}. The method is manifestly covariant, it does not require splitting in space and time.

Consider the involutive closure\footnote{By involutive closure, we mean completion of the system by the generating set of all the admissible lower order consequences. Once the system includes both third and second-order equations, we add the third-order consequences. It is the involutive closure that allows one to count the number of degrees of freedom \cite{Kaparulin:2012px} } of system (\ref{eq_nordstrom_conditional})
\begin{equation}\label{Nordstrom-closure}
    \begin{split}
        L_{\alpha \beta \gamma} \equiv& \partial_{\beta}\partial_{\gamma}\partial^{\lambda}h_{\alpha \lambda}-\partial_{\alpha}\partial_{\beta}\partial^{\lambda}h_{\gamma \lambda}-\partial_{\gamma}\Box h_{\alpha \beta}+ \partial_{\alpha} \Box h_{\beta \gamma}=0\,,\\
        L_{\alpha}\equiv&\partial_{\alpha}N=\partial_{\alpha}\partial^{\mu}\partial^{\nu}h_{\mu \nu}- \Box\partial_{\alpha} h=0\,,\\
         N \equiv&\partial^{\mu}\partial^{\nu}h_{\mu \nu}- \Box h-4\Lambda=0\,.
    \end{split}
\end{equation}
$L_{\alpha \beta \gamma}$ has the hook symmetry in antisymmetric basis,
\begin{equation}
    L_{\alpha \beta \gamma}=-L_{\gamma \beta \alpha}, \quad L_{[\alpha \beta \gamma]}=0\,,
\end{equation}
where brackets mean the antisymmetrization of corresponding indices.
There exist identities between equations (\ref{Nordstrom-closure}),
\begin{equation}
\begin{split}
\label{nordstrom_conditional_identities}
    T^{(1)}_{\mu \nu \rho \omega} \equiv &\,\partial_{\omega}L_{\mu \nu \rho}+\partial_{\mu} L_{\rho \nu \omega}+\partial_{\rho}L_{\omega \nu \mu}=0\,,\\
    T^{(1)}_{\alpha} \equiv&\, \eta^{ \beta \gamma}L_{\beta \gamma \alpha}-\partial_{\alpha}N=0\,,\\
    \widetilde{T}^{(1)}_{\alpha}\equiv &\,L_{\alpha}-\partial_{\alpha}N=0\,.
\end{split}
\end{equation}
There are identities between these identities
\begin{equation}
\label{nordstrom_conditional_identities2}
    T^{(2)}_{\alpha \beta \gamma \delta \omega}\equiv \partial_{\alpha} T^{(1)}_{\gamma \beta \omega \delta}-\partial_{\gamma}T^{(1)}_{\alpha \beta \omega \delta}+\partial_{\delta}T^{(1)}_{\alpha \beta \omega \gamma}-\partial_{\omega}T^{(1)}_{\alpha \beta \delta \gamma}=0\,.
\end{equation}
Equations (\ref{Nordstrom-closure}) have a gauge symmetry, linearized diffeomorphisms,
\begin{equation}
\label{nordstrom_conditional_gauge}
    \delta h^{\mu \nu}=\partial^{\mu}\xi^{\nu}+\partial^{\nu}\xi^{\mu}\,.
\end{equation}
We have explicitly verified by the Macaulay2 package \cite{M2} that (\ref{nordstrom_conditional_identities}), (\ref{nordstrom_conditional_identities2}), and (\ref{nordstrom_conditional_gauge}) are all the independent identities and gauge symmetries of system (\ref{Nordstrom-closure}).
Knowing all identities and gauge symmetries of the involutive system, the number of degrees of freedom in phase space (twice the number of physical polarizations) is calculated by the following formula \cite{Kaparulin:2012px}:
\begin{equation}\label{NDoF2}
\displaystyle \mathcal{N}=\sum\limits_{n}n\Big(t_n-\sum\limits_{m}(-1)^m\big(l_n^m+r_n^m\big)\Big)\,,
\end{equation}
where $t_n$ is the number of equations of order $n$, $l_n^m$ is the number of gauge identities of order $n$ and order of reducibility $m$, $r_n^m$ is the number of gauge symmetries of order $n$ and order of reducibility $m$. For the involutive system (\ref{Nordstrom-closure}) these  non-zero numbers are
\begin{equation}
    t_2=1, \quad t_3=24, \quad r^0_1=4, \quad l^0_3=8 ,\quad l^0_4=15, \quad l^1_5=4\,.
\end{equation}
Using the formula (\ref{NDoF2}), we see the number of degrees of freedom $\mathcal{N}=6$, while for the Einstein equations $\mathcal{N}=4$. It means that this system has two more arbitrary functions in the general solution than a solution of the Einstein equations.

Let us clarify the meaning of additional degrees of freedom in these equations. To do that, we introduce an auxiliary symmetric tensor field $S_{\alpha \beta}$ such that on-shell reduces to the Einstein tensor. Making use of this tensor, we equivalently rewrite system (\ref{eq_nordstrom_conditional}) as follows
\begin{equation}
\label{eq_nordstrom_conditional_S}
    \begin{split}
        &L_{\alpha \beta \gamma} \equiv \partial_{\gamma}S_{\alpha \beta}-\partial_{\alpha}S_{\beta \gamma}-\frac{1}{3}\eta_{\alpha \beta}\eta^{\mu \nu}\partial_{\gamma}S_{\mu \nu}+\frac{1}{3}\eta_{\beta \gamma}\eta^{\mu \nu}\partial_{\alpha}S_{\mu \nu}=0\,,\\
        &N\equiv \eta^{\mu \nu}S_{\mu \nu}=0\,,\\
        &E_{\mu \nu}\equiv \frac{\delta S_{\text{EH}}}{\delta h^{\mu \nu}}-S_{\mu \nu}=0\,.
    \end{split}
\end{equation}
Given these equations, $S_{\alpha \beta}$ can be reinterpreted as a stress-energy tensor since it is the right-hand side (RHS) of the Einstein equations. This tensor is on-shell  divergenceless, since $\eta^{\beta \gamma}L_{\beta \gamma \alpha}=-\partial^{\beta}S_{\beta \alpha}$. Consider the equation in the first line. Let us seek the solution to this equation in the form of a Taylor expansion in the neighborhood of the point $x_0$,
\begin{equation}
\label{eq_nordstrom_conditional_L_ansatz}
    S_{\alpha \beta}=\sum_{n=0}^\infty \frac{1}{n!} A_{\alpha \beta \lambda_1 \dots \lambda_n}(x-x_0)^{\lambda_1}\dots(x-x_0)^{\lambda_n}\,.
\end{equation}
where $A_{\alpha \beta \lambda_1 \dots \lambda_n}$ is the arbitrary tensor being symmetric in the first two indices. Substituting the expansion (\ref{eq_nordstrom_conditional_L_ansatz}) into equations $L_{\alpha \beta \gamma}$  (\ref{eq_nordstrom_conditional_S}) we arrive at the following solution
\begin{equation}
    A_{\alpha \beta \lambda_1 \dots \lambda_n}=S_{\alpha \beta \lambda_1 \dots \lambda_n}-\eta_{\alpha \beta}\eta^{\mu \nu}S_{\mu \nu \lambda_1 \dots \lambda_n}\,,
\end{equation}
where $S_{\alpha \beta \lambda_1 \dots \lambda_n}$ is the totally symmetric tensor. So we see that the general solution to the equations $L_{\alpha \beta \gamma}$ read
\begin{equation}
\label{S_general_sol}
    S_{\alpha \beta}=(\partial_{\alpha}\partial_{\beta}-\eta_{\alpha \beta} \Box)\varphi\,,
\end{equation}
where $\varphi$ is the arbitrary scalar field
\begin{equation}
    \varphi=  \sum_{n=0}^\infty \frac{1}{n!} S_{ \lambda_1 \dots \lambda_n}(x-x_0)^{\lambda_1}\dots(x-x_0)^{\lambda_n}\,.
\end{equation}
Substituting this solution into the Nordstr\"om equation $N=0$ (\ref{eq_nordstrom_conditional_S}), we arrive at the d'Alembert equation for the scalar field $\Box \varphi=0$. Thus the system of equations (\ref{eq_nordstrom_conditional_S}), where all the equations containing $S_{\alpha \beta}$ only are solved, reads
\begin{equation}
\label{nordstrom_and_conditional}
    \frac{\delta S_{\text{EH}}}{\delta h^{\mu \nu}}-\partial_{\mu}\partial_{\nu}\varphi=0\,, \quad \Box \varphi=0\,.
\end{equation}
This system of equations describes linearized Einstein gravity with a massless scalar field. The two additional degrees of freedom appear because of the presence of a scalar field, which obeys the d'Alembert equation. We note that equations (\ref{nordstrom_and_conditional}) can be obtained from the action
\begin{equation}
\label{S_phi_Lagrange}
    S_{\varphi} = \int d^4x (\mathcal{L}_{\text{EH}}-\varphi N)\,,
\end{equation}
where $N$ is the left-hand side (LHS) of the Nordstr\"om equation (\ref{nordstrom_linearized}) and $\varphi$ is the Lagrange multiplier for Nordstr\"om equation. Equations obtained by varying the action in the class of metrics restricted by the linearized Nordstr\"om equation and Nordstr\"om equation itself, and equations obtained from the action with Lagrange multiplier (\ref{S_phi_Lagrange}) coincide. This is quite natural because, for the pure gauge equations (\ref{eq_general}), the Lagrange multiplier method is equivalent to the partially Lagrangian system (\ref{eq_general}),(\ref{eq_conditional_extremum_general}) as explained in subsection 2.2. The Nordstr\"om equation, being considered irrespectively to other equations on metrics, has no physical degrees of freedom --- it is a pure gauge system \cite{Abakumova:2022eoo}.

Consider the system (\ref{eq_nordstrom_conditional_S}) without  Nordstr\"om equation $N=0$. Substituting general solution (\ref{S_general_sol}) in $E_{\mu \nu}$ (\ref{eq_nordstrom_conditional_S}) we arrive at the following equations
\begin{equation}
\label{nordstrom_eq_phi}
    E_{\mu \nu} \equiv \frac{\delta S_{\text{EH}}}{\delta h^{\mu \nu}}-(\partial_{\mu}\partial_{\nu}-\eta_{\mu \nu} \Box)\varphi=0\,.
\end{equation}
They admit a consequence
\begin{equation}
\label{E_consequence}
    \eta^{\mu \nu} E_{\mu \nu} \equiv -N+3\Box \varphi = 0\,.
\end{equation}
Equations (\ref{nordstrom_eq_phi}), (\ref{E_consequence}) have a conformal gauge symmetry
\begin{equation}
\label{gauge_sym_withoutNordstrom}
    \delta h^{\mu \nu}=-\eta^{\mu \nu} \epsilon\,, \quad \delta \varphi=\epsilon\,.
\end{equation}
They can be derived from the action
\begin{equation}\label{EH-phi-action}
    S_{\text{conformal}}=\int d^4x[\mathcal{L}_{\text{EH}}-\frac{3}{2}\partial_{\mu}\varphi\partial^{\mu}\varphi-\varphi(\partial^{\mu}\partial^{\nu}h_{\mu \nu}-\Box h-4\Lambda)]\,.
\end{equation}
This is the linearized action of a conformal scalar field coupled with gravity \cite{Birrell:1982ix}. Let us make change of the field variables: $h_{\mu \nu}\rightarrow h_{\mu \nu}-\eta_{\mu \nu}\varphi$. After the change, $\varphi$  drops out, and (\ref{EH-phi-action}) reduces to the Einstein-Hilbert action. This means that the theory described by equations (\ref{nordstrom_eq_phi}) is equivalent to the linearized Einstein gravity. This fact is a special case of the well-known equivalence between various gravity actions with a scalar field \cite{Bettoni:2013diz}.

Let us express  $S_{\alpha \beta}$ in terms of the linearized Ricci tensor and scalar curvature in the equations $E_{\mu \nu}$ of (\ref{eq_nordstrom_conditional_S}). Substituting that into  $L_{\alpha \beta \gamma}$, we arrive at the expression
\begin{equation}
    L_{\alpha \beta \gamma}=\partial_{\gamma}R^{(\ell)}_{\alpha \beta}-\partial_{\alpha}R^{(\ell)}_{\beta \gamma}+\frac{1}{6}(\eta_{\beta \gamma}\partial_{\alpha}R^{(\ell)}-\eta_{\alpha \beta}\partial_{\gamma}R^{(\ell)})\,,
\end{equation}
where $(\ell)$ denotes the linearized tensor. One can see that $L_{\alpha \beta \gamma}$, being the LHS of the partially Lagrangian equations (\ref{eq_nordstrom_conditional}), is the linearized Cotton tensor. The Cotton tensor is defined as follows
\begin{equation}\label{Cotton-definition}
    C_{\beta \alpha \gamma}=\nabla_{\gamma}R_{\alpha \beta}-\nabla_{\alpha}R_{\beta \gamma}+\frac{1}{6}(g_{\beta \gamma}\nabla_{\alpha}R-g_{\alpha \beta}\nabla_{\gamma}R)\,.
\end{equation}
As we see, the extrema of the Einstein-Hilbert action in the class of trajectories restricted by Nordstr\"om equation (\ref{eq_nordstrom_conditional}) obey at the linear level equations  $C_{\alpha \beta \gamma}=0$ known as Cotton gravity \cite{Harada:2021bte}. As we demonstrate above, Cotton gravity is equivalent to the Einstein theory at the linear level.

\noindent
Let us discuss the possibility of extending the conclusion about the equivalence of the Cotton gravity equations and Einstein ones beyond the linear level. Gauge symmetry is obviously extended, as the linearized diffeomorphism can be replaced by the complete one.  As the Cotton gravity equations are not Lagrangian, there is no pairing between gauge symmetries and gauge identities. Consistent deformation of the free theory implies that the number and order of the identities have to remain the same as at the linear level. Consider the minimal covariantization of identities $T^{(1)}_{\alpha \beta \gamma \delta}$ in (\ref{nordstrom_conditional_identities})
\begin{equation}
             T^{(1)}_{\alpha \beta \gamma \delta} \equiv \nabla_{\delta}C_{\beta \alpha \gamma}+\nabla_{\alpha} C_{\beta \gamma \delta}+\nabla_{\gamma}C_{\beta \delta \alpha}\,.
\end{equation}
Substituting the definition of Cotton tensor, one can find the identity (see in \cite{Garcia:2003bw})
\begin{equation}
\label{cotton_ricci_identity}
    \nabla_{\delta}C_{\beta \alpha \gamma}+\nabla_{\alpha} C_{\beta \gamma \delta}+\nabla_{\gamma}C_{\beta \delta \alpha}=R_{\delta \gamma \beta \lambda}R^{\lambda}{}_{\alpha}+R_{\alpha \delta \beta \lambda}R^{\lambda}{}_{\gamma}+R_{\gamma \alpha \beta \lambda}R^{\lambda}{}_{\delta}\,.
\end{equation}
So, the minimal covariantization of the gauge identity in the first line of relations (\ref{nordstrom_conditional_identities}) is broken by the quadratic terms in the curvature tensor. We introduce the following notation for minimal covariantization of linearized equations and gauge identity generators
\begin{equation}
    \begin{split}
        \rho^{(0)\alpha^{'} \beta^{'} \gamma^{'}}_{\alpha \beta \gamma \delta} =&\frac{1}{3}[ (\delta^{\alpha^{'}}_{\beta} \delta^{\beta^{'}}_{\alpha} \delta^{\gamma^{'}}_{\gamma}
+\delta^{\alpha^{'}}_{\alpha} \delta^{\beta^{'}}_{\beta} \delta^{\gamma^{'}}_{\gamma}
-\delta^{\alpha^{'}}_{\beta} \delta^{\beta^{'}}_{\gamma} \delta^{\gamma^{'}}_{\alpha}
-\delta^{\alpha^{'}}_{\gamma} \delta^{\beta^{'}}_{\beta} \delta^{\gamma^{'}}_{\alpha}) \nabla_{\delta}
+(\delta^{\alpha^{'}}_{\beta} \delta^{\beta^{'}}_{\gamma} \delta^{\gamma^{'}}_{\delta}
+\delta^{\alpha^{'}}_{\gamma} \delta^{\beta^{'}}_{\beta} \delta^{\gamma^{'}}_{\delta}
\\
&-\delta^{\alpha^{'}}_{\beta} \delta^{\beta^{'}}_{\delta} \delta^{\gamma^{'}}_{\gamma}
-\delta^{\alpha^{'}}_{\delta} \delta^{\beta^{'}}_{\beta} \delta^{\gamma^{'}}_{\gamma}) \nabla_{\alpha}
+(\delta^{\alpha^{'}}_{\beta} \delta^{\beta^{'}}_{\delta} \delta^{\gamma^{'}}_{\alpha}
+\delta^{\alpha^{'}}_{\delta} \delta^{\beta^{'}}_{\beta} \delta^{\gamma^{'}}_{\alpha}
-\delta^{\alpha^{'}}_{\beta} \delta^{\beta^{'}}_{\alpha} \delta^{\gamma^{'}}_{\delta}
-\delta^{\alpha^{'}}_{\alpha} \delta^{\beta^{'}}_{\beta} \delta^{\gamma^{'}}_{\delta}) \nabla_{\gamma}]\,,\\
L^{(0)}_{\alpha \beta \gamma}=&C_{\beta \alpha \gamma}, \quad \rho^{(0)\alpha^{'} \beta^{'} \gamma^{'}}_{\alpha \beta \gamma \delta} L^{(0)}_{\alpha^{'} \beta^{'} \gamma^{'}}=T^{(1)}_{\alpha \beta \gamma \delta} = \nabla_{\delta}C_{\beta \alpha \gamma}+\nabla_{\alpha} C_{\beta \gamma \delta}+\nabla_{\gamma}C_{\beta \delta \alpha}\,.
    \end{split}
\end{equation}
The gauge identity generator $\rho^{(0)\alpha^{'} \beta^{'} \gamma^{'}}_{\alpha \beta \gamma \delta}$ is of the zero order in curvature, equation $L^{(0)}_{\alpha \beta \gamma}$ is of the first order. Let us try to deform gauge identity generators and equations to make the RHS of relation (\ref{cotton_ricci_identity}) vanish, in such a manner that the linearized part of equations and identities remain unchanged. We also have to take into account that the deformation should not increase the orders of the identities and the equations to preserve the number of DoFs. These conditions imply the following most general structure of deformation in the next order of the curvature tensor:
\begin{equation}
    \rho = \rho^{(0)} + \delta \delta \delta R \nabla + \cdots, \quad L = L^{(0)} + R \nabla R + g R \nabla R + \cdots\,,
\end{equation}
where $\cdots$ denote the terms of higher order in the Riemann tensor. Here, we do not explicitly write the indices. The gauge identity after deformation has the following structure:
\begin{equation}
    \rho L = T^{(1)} + \nabla R \nabla R +g\nabla R \nabla R+ R \nabla \nabla R + g R \nabla \nabla R +\cdots\,.
\end{equation}
Extra terms arising from deformation cannot be canceled out by the terms in $T^{(1)}$. It means that the gauge identity $T^{(1)}$ (\ref{nordstrom_conditional_identities}), being linear in perturbation of the metric, cannot be consistently deformed to the full non-linear theory. So, we see the obstruction to consistent deformation of the linearized Cotton gravity to the nonlinear level. This does not necessarily mean inconsistency of the Cotton gravity at the nonlinear level, but the equivalence with GR remains questionable, while at the linear approximation, they are equivalent, as we have shown above.  From the examples of special solutions, the distinctions between Cotton gravity and Einstein equations are mentioned in the recent literature, see \cite{Clement:2023tyx} and references therein. Here we see a more fundamental source of non-equivalence.

\section{Concluding remarks }

Let us briefly summarize the results and discuss open questions.

In this article, we consider the problem of the conditional extrema for the action in the class of trajectories restricted by a system of differential equations (\ref{eq_general}). Our method implies finding, at first, the infinitesimal gauge symmetry of these equations considered irrespectively to the action functional. As the second step, we replace the unfree variation of the action with the gauge variation with respect to the infinitesimal gauge transformation (\ref{gauge_transformaion_general}), (\ref{gauge_transformation_def}) of the equations (\ref{eq_general}) restricting the trajectories.
Since the unfree variation must vanish for conditional extrema in the class of fields restricted by equations (\ref{eq_general}), we arrive at the system of equations (\ref{eq_conditional_extremum_general}) for conditionally critical trajectories of the action. These equations involve only original fields and do not involve Lagrange multipliers. We demonstrate that the system of Lagrangian equations for the action with Lagrange multipliers (\ref{S-lambda}) would have the same solutions for the original fields as the partially Lagrangian system (\ref{eq_general}), (\ref{eq_conditional_extremum_general}). The Lagrange multipliers bring in extra degrees of freedom, in general, except for some special cases, including algebraic systems (\ref{eq_general}). The additional degrees of freedom, being brought by Lagrange multipliers, are unrelated to the original problem of conditional extremum for the action, and they can spoil the positivity of energy. The partially Lagrangian equations (\ref{eq_general}), (\ref{eq_conditional_extremum_general}) for conditional extrema of the action can be systematically brought to the Hamiltonian form, though the Poisson bi-vector is degenerate, in general. The exception is the case when the equations (\ref{eq_general}), which restrict trajectories, describe a pure gauge system if they are considered irrespectively to the action.  Since the partially Lagrangian equations (\ref{eq_general}), (\ref{eq_conditional_extremum_general}) admit Hamiltonian formulation, these systems can be quantized, in principle.

We illustrate the general method with two examples intended to demonstrate the distinctions of the partially Lagrangian dynamics both from the usual variational systems and from the equations with Lagrange multipliers.  The first example is a mechanical model of a non-relativistic particle on the plane in a central potential with the usual action (\ref{dMdt_action}) and the class of trajectories restricted by equation (\ref{dMdt}) of conservation of angular momentum. If the restriction was not imposed, the system would have 4 degrees of freedom by the phase space count. The partially Lagrangian system that defines conditionally critical trajectories includes one second-order equation (\ref{dMdt}), and one equation of the third order (\ref{dMdt_condtitional_equation}). So, there are 5 degrees of freedom.
For the action with Lagrange multiplier (\ref{dMdtS-lambda}), the dynamics has 6 degrees of freedom.
Inclusion of the multiplier makes the energy always unbounded,
with any initial data for the original variables $r,\phi$.
The partially Lagrangian equations (\ref{dMdt}), (\ref{dMdt_condtitional_equation}) govern the dynamics
that admits, besides conserved energy and momentum,  the extra conserved quantity -- precession parameter $K$ -- defined by relation  (\ref{K}). Solutions with $K=0$ reproduce the usual Lagrangian dynamics of the particle in the central potential field. For $K\neq 0$, the equations remain integrable. As we see for the Kepler potential $U=-\frac{\alpha}{r}$, with $K\neq 0$, the particle dynamics may be either bounded motion with precession, unbounded, or it can fall into the center, depending on the initial data.

As the second example of a partially Lagrangian system, we consider the problem of conditional extremum of the linearized General Relativity action (\ref{EH_linearized}) in the class of metrics restricted by linearized Nordstr\"om equation (\ref{nordstrom_linearized}).  Full gauge symmetry of the Nordstr\"om equation is known at linear level \cite{Abakumova:2022eoo}, see (\ref{norstrom-gauge-sym}).  Since the gauge symmetry of the Nordstr\"om equation is parameterized by the third rank tensor with the hook symmetry, the equations for conditional extrema of the linearized Einstein-Hilbert action have the same symmetry. They turn out to be linearized equations of Cotton gravity. This is quite natural since the Cotton tensor has the hook symmetry and involves the third-order derivatives of the metrics. As we demonstrate, the Cotton equations (\ref{eq_nordstrom_conditional}), being considered independently from the Nordstr\"om equation (\ref{nordstrom_linearized}), are equivalent to the Einstein equations at the linear level. We also identify obstructions to extending this conclusion beyond the linear level. The complete partially Lagrangian system (\ref{eq_nordstrom_conditional}) includes Cotton and Nordstr\"om equations. This system has 6 local degrees of freedom by the phase space count.  We demonstrate that these partially Lagrangian third-order equations for metrics are equivalent to the linearization of the Lagrangian system of the massless scalar field coupled to the metrics (\ref{S_phi_Lagrange}). So, imposing the Nordstr\"om equation (\ref{nordstrom_linearized}) as the condition that restricts the class of varying fields, we bring the extra degree of freedom to the pure metric theory with the Einstein-Hilbert action. This degree of freedom turns out to be a scalar.

As can be seen from the examples, the restrictions imposed by differential equations on the class of varying trajectories can make the dynamics that follow from the variational principle with the same action functional significantly more diverse.
Upon restrictions imposed on the trajectories by the differential equations, all the dynamics of the Lagrangian system of the corresponding action survive as special solutions, while more solutions become admissible. We see the extra degrees of freedom arise, which allow one to describe more phenomena, including new types of evolutions and more conserved quantities, without the inclusion of new fields.

Let us now discuss the open issues and further perspectives.

First, in section 3, we demonstrate that partially Lagrangian systems can be brought to Hamiltonian form (\ref{hamiltonian_gen}), (\ref{eq_hamiltonian_drift_general}) with the drift (\ref{Drift-gen}) such that differentiates the Poisson brackets (\ref{brackets_general}), (\ref{Drift-Leibnitz-gen}). This form of dynamics can be quantized \cite{Lyakhovich:2004xd}.
To come to this Hamiltonian formalism, we proceed from the first-order formulation (\ref{normal_form_eq}), (\ref{L_x_u}) of the conditional extrema problem, imposing an auxiliary requirement that the Hessian is non-degenerate (\ref{hessian}). The partially Lagrangian equations for conditional extrema are well defined with a degenerate Hessian, but the construction of the Hamiltonian formalism has to be reconsidered for this case. One would have constraints on the phase space and possibly gauge symmetry of dynamics. As the equations are not variational, the constraints are not paired with gauge symmetry, see in \cite{Lyakhovich:2008hu}. The extension of the Hamiltonian formalism to the case of degenerate Hessian would probably lead to the Hamiltonian equations with the weak Poisson brackets such that obey Jacobi identity for gauge invariants (not for all phase space functions) and modulo constraints. This is sufficient to retain all the significant properties of Hamiltonian formalism, and it allows for quantization \cite{Lyakhovich:2004xd}. One more aspect left aside when the Hamiltonian formalism is constructed in Section 3 is the option to have an over-complete generating set of the vectors in the characteristic distribution $\bar{\mathcal{Z}}_V$ (\ref{Z-bar}) that defines gauge symmetry of the equations (\ref{normal_form_eq}) restricting the trajectories in the first order formalism. This would mean that the gauge symmetry of the equations (\ref{normal_form_eq}) is reducible, and this should be addressed in the construction. The most natural way to address that seems to impose the extra constraints on the momenta $p_{\bar{\alpha}}$ being generated by the null vectors for $Z_{\bar{\alpha}}$. Models of this type may naturally arise in the field theory, as we see from the second example of section 4.

The second remark is that for partially Lagrangian systems (\ref{eq_general}), (\ref{eq_conditional_extremum_general}), whose dynamics follow from the problem of conditional extrema for the action functional, the first and the second Noether theorems are not directly applicable. These theorems proceed from the unconditional least action principle, so the variation is unrestricted. The inclusion of the conditions, being differential equations, with Lagrange multipliers, into the action would bring, in general, extra degrees of freedom to the theory. Corresponding Noether conserved quantities and identities would involve the non-physical degrees of freedom brought by Lagrange multipliers, so this is irrelevant to the original dynamics defined by the conditional extrema problem. The extension of the Noether theorems to the partially Lagrangian systems seems to be related to the structure of the Lagrange anchor introduced in the article \cite{Kazinski:2005eb}.
If not necessarily Lagrangian equations of motion admit the Lagrange anchor, the latter connects conserved quantities with rigid symmetries \cite{Kaparulin:2010ab}, and the Noether identities are connected to gauge symmetries \cite{Kazinski:2005eb}. We expect that partially Lagrangian systems with equations of motion (\ref{eq_general}), (\ref{eq_conditional_extremum_general}) should admit a Lagrange anchor naturally defined by the operators $\hat{R}_\alpha$ (\ref{gauge_transformaion_general}).

Finally, let us make the overall concluding remark. The general setup of partially Lagrangian dynamics retains all the most essential features of field theories following from the unconditional least action principle and provides tools for describing a wider class of phenomena compared to pure variational systems.

\vspace{2 mm}

\subsection*{Acknowledgments.}
The work is supported by the research project FSWM-2025-0007 of the Ministry of Science and  Higher Education of Russian Federation.

The authors thank D.S.Kaparulin for collaboration on the first example at the early stage of the work.
We are also grateful to A.A.~Sharapov for the valuable discussions.



\begin{thebibliography}{111}

\bibitem{Crnkovic}
Crnkovic, C. and Witten, E., 1987. ``Covariant description of canonical formalism in geometrical theories''. In: ``Three hundred years of gravitation", ed. by S.W. Hawking, W. Israel (Cambridge University Press,
Cambridge, 1987) pp.676-684.


\bibitem{Havkine} I.Khavkine, ``Presymplectic current and the inverse problem of the calculus of variations.''
J. Math. Phys. 54, 111502 (2013)



\bibitem{Sharapov-tricomplex} A.A. Sharapov, ``Variational Tricomplex, Global Symmetries and Conservation Laws of Gauge
Systems'', SIGMA, 2016, \textbf{12}, 098


\bibitem{Alkalaev:2013hta}
K.~B.~Alkalaev and M.~Grigoriev,
``Frame-like Lagrangians and presymplectic AKSZ-type sigma models,''
Int. J. Mod. Phys. A \textbf{29} (2014) no.18, 1450103
doi:10.1142/S0217751X14501036
[arXiv:1312.5296 [hep-th]].


\bibitem{Sharapov-presymplectic} A.A. Sharapov, ``On presymplectic structures for massless higher-spin fields'',
Eur. Phys. J. C (2016) 76:305

\bibitem{Dneprov:2024cvt}
I.~Dneprov, M.~Grigoriev and V.~Gritzaenko,
``Presymplectic minimal models of local gauge theories,''
J. Phys. A \textbf{57} (2024) no.33, 335402
doi:10.1088/1751-8121/ad65a3
[arXiv:2402.03240 [hep-th]].

\bibitem{Grigoriev:2021wgw}
M.~Grigoriev and V.~Gritzaenko,
``Presymplectic structures and intrinsic Lagrangians for massive fields,''
Nucl. Phys. B \textbf{975} (2022), 115686
doi:10.1016/j.nuclphysb.2022.115686
[arXiv:2109.05596 [hep-th]].


\bibitem{Lyakhovich:2004xd}
S.~L.~Lyakhovich and A.~A.~Sharapov,
``BRST theory without Hamiltonian and Lagrangian,''
JHEP \textbf{03} (2005), 011.


\bibitem{Kazinski:2005eb}
P.~O.~Kazinski, S.~L.~Lyakhovich and A.~A.~Sharapov,
``Lagrange structure and quantization,''
JHEP \textbf{07} (2005), 076.


\bibitem{Lyakhovich:2006sc}
S.~L.~Lyakhovich and A.~A.~Sharapov,
``Quantizing non-Lagrangian gauge theories: An Augmentation method,''
JHEP \textbf{01}, 047 (2007)
doi:10.1088/1126-6708/2007/01/047
[arXiv:hep-th/0612086 [hep-th]].


\bibitem{Lyakhovich:2007cw}
S.~L.~Lyakhovich and A.~A.~Sharapov,
``Quantization of Donaldson-Uhlenbeck-Yau theory,''
Phys. Lett. B \textbf{656}, 265-271 (2007)
doi:10.1016/j.physletb.2007.09.029
[arXiv:0705.1871 [hep-th]].


\bibitem{Kaparulin:2010ab}
D.~S.~Kaparulin, S.~L.~Lyakhovich and A.~A.~Sharapov,
``Rigid Symmetries and Conservation Laws in Non-Lagrangian Field Theory,''
J. Math. Phys. \textbf{51}, 082902 (2010)
doi:10.1063/1.3459942
[arXiv:1001.0091 [math-ph]]

\bibitem{Gelfand}
Gelfand, I.M. and Silverman, R.A., 2000. Calculus of variations. Courier Corporation.
\bibitem{Henneaux:1992ig}
M.~Henneaux and C.~Teitelboim,
``Quantization of gauge systems'' (Princeton U.P., NJ, 1992).

\bibitem{Dirac} P.A.M. Dirac, Lectures on Quantum Mechanics, Yeshiva University, New York: Academic Press, 1967.

\bibitem{Lyakhovich:2008hu}
S.~L.~Lyakhovich and A.~A.~Sharapov,
``Normal forms and gauge symmetries of local dynamics,''
J. Math. Phys. \textbf{50} (2009), 083510.

\bibitem{Agrachev}
Agrachev, A. \& Sachkov, Y. Control Theory from the Geometric Viewpoint. {\em Encyclopaedia Of Mathematical Sciences}. (2004)

\bibitem{Goldstein}  H. Goldstein, Classical Mechanics (2nd Edition), Addison-Wesley, 1980.

\bibitem{Abakumova:2022eoo}
V.~Abakumova, D.~Frolovsky, H.~C.~Herbig and S.~Lyakhovich,
``Gauge symmetry of linearised Nordstr\"om gravity and the dual spin two field theory,''
Eur. Phys. J. C \textbf{82} (2022) no.9, 780.

\bibitem{Kaparulin:2012px}
D.~S.~Kaparulin, S.~L.~Lyakhovich and A.~A.~Sharapov,
``Consistent interactions and involution,''
JHEP \textbf{01} (2013), 097.


\bibitem{M2}
D.~R.~Grayson and Michael~E.~Stillman,
``Macaulay2, a software system for research in algebraic geometry'',
Available at \url{http://www.math.uiuc.edu/Macaulay2/}.


\bibitem{Birrell:1982ix}
N.~D.~Birrell and P.~C.~W.~Davies,
``Quantum Fields in Curved Space,''
Cambridge University Press, 1982.


\bibitem{Bettoni:2013diz}
D.~Bettoni and S.~Liberati,
``Disformal invariance of second order scalar-tensor theories: Framing the Horndeski action,''
Phys. Rev. D \textbf{88}, 084020 (2013).


\bibitem{Harada:2021bte}
J.~Harada,
``Emergence of the Cotton tensor for describing gravity,''
Phys. Rev. D \textbf{103} (2021) no.12, L121502


\bibitem{Garcia:2003bw}
A.~Garcia, F.~W.~Hehl, C.~Heinicke and A.~Macias,
``The Cotton tensor in Riemannian space-times,''
Class. Quant. Grav. \textbf{21} (2004), 1099-1118

\bibitem{Clement:2023tyx}
G.~Cl\'ement and K.~Nouicer,
``Cotton gravity is not predictive,''
Phys. Lett. B \textbf{856}, 138947 (2024)
doi:10.1016/j.physletb.2024.138947
[arXiv:2312.17662 [gr-qc]].


\end{thebibliography}
\end{document}